# Lone-Pair Stereoactivity Overrides Mass Scaling of Thermal Expansion and Heat Conduction in Non-van der Waals Layered Bismuth Oxychalcogenides

Linzhen Li[1,2,†], Guangwu Zhang[2,†], Ziyan Qian[2,†], Wei Jing[2], Yang Lu[2], Jiaqi Gu[2], Kaifeng Huang[2], Yanguang Zhou[2], and Qiye Zheng[2,*]

[1]School of Energy Engineering, Zhejiang University, Hangzhou 310027, China

[2]Department of Mechanical and Aerospace Engineering, The Hong Kong University of Science and Technology, Hong Kong SAR, China

[†]These authors contributed equally.

[*]Corresponding author: qiyezheng@ust.hk

**Abstract**

Heavier chalcogens are generally expected to soften isostructural semiconductor lattices, raising thermal expansion and lowering thermal conductivity. Layered bismuth oxychalcogenides, $Bi_2O_2X$ ($X = S, Se, Te$), are mixed-anion crystals whose covalent $[Bi_2O_2]^{2+}$ slabs and ionic $X^{2-}$ sheets are bound electrostatically rather than by van der Waals forces and whose $Bi^{3+}$ $6s^2$ lone pairs can become stereochemically active. Despite the rapid development of $Bi_2O_2Se$ for fin and gate-all-around transistors, its fundamental thermal expansion and heat transport mechanisms remain poorly understood. Here, we show that lone-pair stereoactivity, not chalcogen mass, governs both properties. Path-integral molecular dynamics shows that the volumetric thermal expansion coefficient at 300 K falls from $8.14\times10^{-5}$ K$^{-1}$ in $Bi_2O_2S$ to 6.72 and $6.61\times10^{-5}$ K$^{-1}$ in $Bi_2O_2Se$ and $Bi_2O_2Te$ despite a 13% larger unit cell in the telluride, whereas substituting the S mass into $Bi_2O_2Se$ changes its value by only 0.6%. Lone-pair-driven Bi off-centering in orthorhombic $Bi_2O_2S$ makes $b$-axis expansion 1.63 times $a$-axis expansion yet leaves in-plane conduction isotropic within 3%. Phonon transport calculations including four-phonon scattering and Wigner coherence predict a conductivity maximum in $Bi_2O_2Se$ (1.83 and 0.95 W m$^{-1}$ K$^{-1}$ in plane and cross plane at 300 K) because $Bi_2O_2S$ is limited by anharmonic scattering and $Bi_2O_2Te$ by slow cross-plane phonons. Thermoreflectance from 300 to 500 K and *in situ* X-ray diffraction on $Bi_2O_2Se$ single crystals agree with these predictions within 6.4%. Phonon Boltzmann transport simulations of 12 nm wide fins, which run hotter than Fourier diffusion predicts, show that $Bi_2O_2Se$ minimizes the hotspot temperature and the thermal-mismatch strain at the gate dielectric.

## 1 Introduction

Layered bismuth oxychalcogenides, $Bi_2O_2X$ ($X = S, Se, Te$), have emerged as a semiconductor platform for post-silicon nanoelectronics and optoelectronics[1–6]. Their Sillén-type lattice alternates covalent, fluorite-like $[Bi_2O_2]^{2+}$ slabs with ionic $X^{2-}$ sheets [Fig. 1(a)]. Because adjacent layers carry opposite formal charges, they are bound by predominantly electrostatic rather than van der Waals (vdW) interactions, and this non-vdW layered motif combines layer-controlled epitaxy[3;7] with air stability, small in-plane effective masses, and high carrier mobilities. $Bi_2O_2Se$, the most mature member, couples a room-temperature electron mobility of about 450 $cm^2 V^{-1} s^{-1}$ with an epitaxial native high-$k$ oxide, $\beta$-$Bi_2SeO_5$, whose atomically sharp interface suppresses short-channel effects in fin and gate-all-around field-effect transistors (FinFETs and GAAFETs)[8–10]. Vertical $Bi_2O_2Se$ fins as thin as one unit cell and gate-all-around channels now deliver on-state currents from 0.83 to above 1 mA $\mu m^{-1}$ [11;14;15]. $Bi_2O_2Te$ nanosheet transistors exploit small effective masses for high-mobility channels, and lattice distortions in atomically thin $Bi_2O_2S$ enable room-temperature ferroelectric switching for nonvolatile memory[4;5;10–15]. Composition-tunable bandgaps further support air-stable visible to near-infrared detection in $Bi_2O_2Se$, silicon-band-edge photodetection in $Bi_2O_2S$, and extended infrared operation in narrow-gap $Bi_2O_2Te$[16–19].

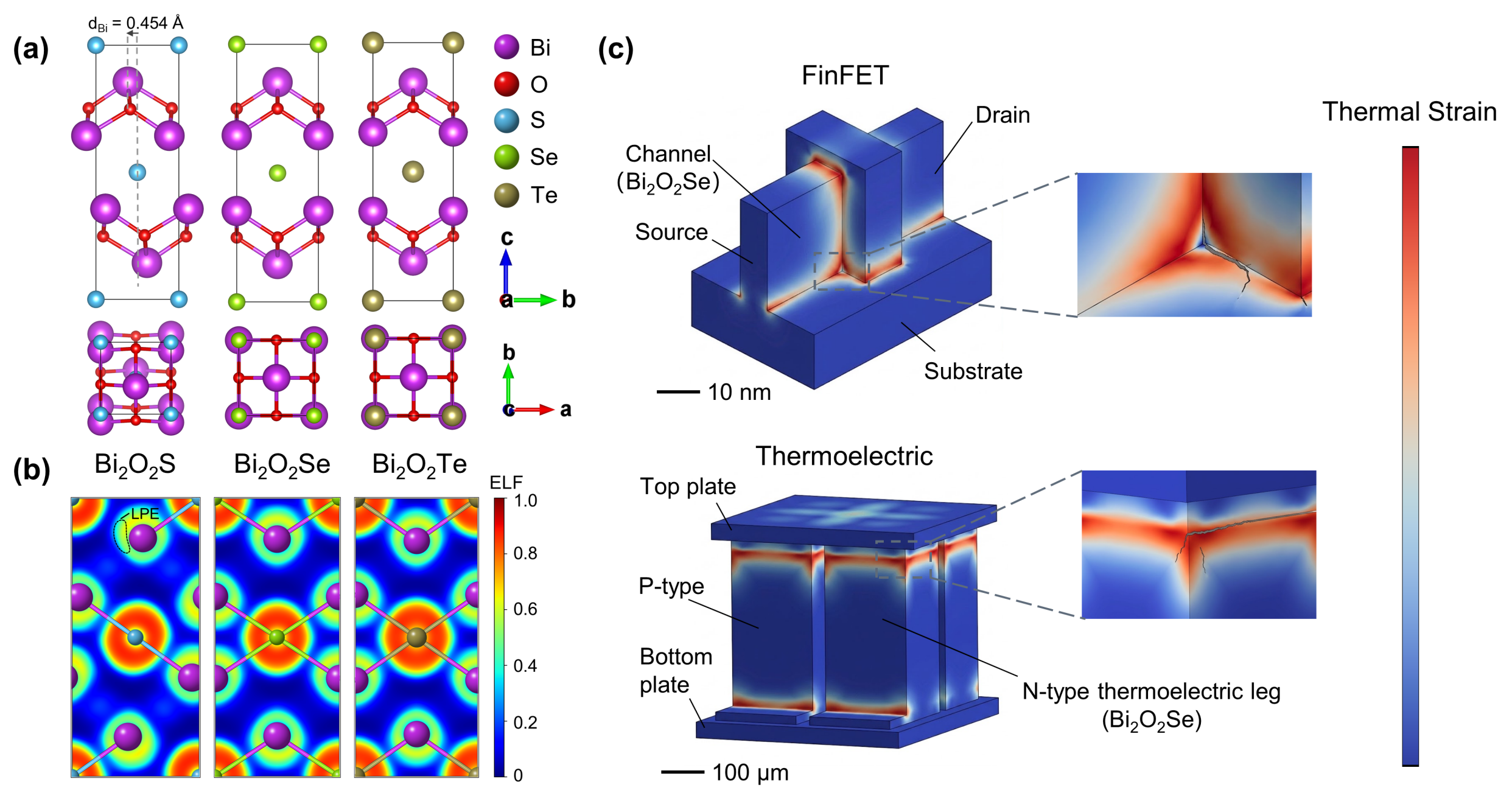

**Figure 1:** (a) Side and top views of $Bi_2O_2S$, $Bi_2O_2Se$, and $Bi_2O_2Te$. Gray wireframes mark conventional unit-cell boundaries. The Bi off-centering along the $b$ axis in $Bi_2O_2S$ is indicated as $d_{Bi} = 0.454$ Å relative to the symmetry-matched $I4/mmm$ reference position. Purple, red, blue, green, and olive spheres denote Bi, O, S, Se, and Te, respectively. (b) Ground-state electron localization function (ELF) maps on the (110) planes of the three compounds, plotted on a common scale from 0 to 1. The dashed outline highlights the asymmetric lone-pair localization adjacent to Bi in $Bi_2O_2S$. (c) Schematic thermal-strain distributions and representative thermomechanical failure scenarios in a FinFET with a $Bi_2O_2Se$ channel (top) and a thermoelectric module with a $Bi_2O_2Se$ n-type leg (bottom). Enlarged callouts illustrate crack initiation and interfacial failure in regions of concentrated thermoelastic strain arising from thermal-expansion mismatch. Scale bars indicate characteristic dimensions, and the color scale represents the thermal-strain magnitude.

These nanoscale architectures confine heat generation and thermal strain to dimensions comparable to or smaller than the mean free paths of heat-carrying phonons. In sub-15 nm channels, phonon–boundary scattering impedes heat removal and invalidates Fourier-based estimates of the hotspot temperature[20–22], while the mismatch between the thermal expansion of the channel and that of the gate dielectric, contacts, and substrate concentrates interfacial stress during self-heating and thermal cycling [Fig. 1(c)][23;24]. Analogous constraints arise in $Bi_2O_2Se$ thermoelectric legs, where a low lattice thermal conductivity ($\kappa_L$) supports figures of merit of $ZT \approx 0.5$–$0.7$ near 800 K but steep gradients and contact mismatch promote microcracking and fatigue[23;24], and in silicon-integrated $Bi_2O_2X$ photodetectors, where thermally induced strain shifts the bandgap and destabilizes the photoresponse[25–30].

Predicting these failure modes requires the full anisotropic tensors of $\kappa_L$ and of the coefficient of thermal expansion (CTE, $\alpha$).

Both tensors are set by how bond stiffness and anharmonicity are partitioned across the layered lattice. In vdW crystals such as graphite, weak interlayer coupling allows flexural membrane modes with negative Grüneisen parameters to contract the basal plane on heating[31]. Mixed-anion compounds instead combine chemically distinct anion sublattices with a built-in hierarchy of covalent and ionic bonds[32]. In $Bi_2O_2Se$, compression shortens the interlayer Bi–Se bonds more than three times faster than the intralayer Bi–O bonds[3], identifying the electrostatic interlayer contact as the compliant element of an otherwise rigid framework. Superimposed on this hierarchy is the $Bi^{3+}$ $6s^2$ lone pair. In the revised lone-pair model, the lone pair becomes stereochemically active when antibonding cation $s$–anion $p$ states mix with empty cation $p$ states upon symmetry lowering[33], which produces asymmetric coordination, soft low-frequency phonons, and near-minimum $\kappa_L$ in lone-pair chalcogenides[34;35]. Because $\alpha_V = (B_T V)^{-1} \sum_\lambda C_{V,\lambda}\gamma_\lambda$, where $B_T$ is the isothermal bulk modulus, $V$ the volume, and $C_{V,\lambda}$ and $\gamma_\lambda$ the heat capacity and Grüneisen parameter of mode $\lambda$, the same lone-pair-softened modes should also control thermal expansion, as the spontaneous-strain and negative-expansion anomalies of lone-pair ferroelectrics such as $PbTiO_3$ illustrate[36].

In $Bi_2O_2X$, lone-pair-related anharmonicity yields large Grüneisen parameters and a broad multiphonon scattering phase space, so that four-phonon (4ph) scattering and wave-like interband tunneling suppress $\kappa_L$ to values comparable to those of SnSe and $Bi_2Te_3$[37–43]. The ELF maps in Fig. 1(b) show that the asymmetry of the lone-pair density varies across the series. Pressure and strain studies support this lone-pair–phonon link. Hydrostatic compression suppresses Bi lone-pair activity in $Bi_2O_2Se$ and reshapes its phonon group velocities until the conduction anisotropy inverts[44], whereas $c$-axis tensile strain in BiCuSeO, which shares the $[Bi_2O_2]^{2+}$ slab, expands the lone-pair lobes, softens optical phonons, and intensifies Umklapp scattering[43]. Chalcogen substitution tunes the same interaction chemically. In tetragonal $Bi_2O_2Se$ and $Bi_2O_2Te$ ($I4/mmm$), the higher-lying and more polarizable Se $4p$ and Te $5p$ states weaken the Bi $6s$–$X$ $np$ antibonding interaction and leave $Bi^{3+}$ centered with $a = b$[44]. In $Bi_2O_2S$, the deeper S $3p$ states strengthen this interaction, activating the lone pair and driving cooperative in-plane Bi off-centering into an orthorhombic $Pnnm$ ground state with $a \neq b$[45], which hydrostatic pressure of a few gigapascals restores to $I4/mmm$[40].

This chemical tuning raises questions that existing data cannot answer. Across most chalcogenide series, heavier anions soften the lattice, so that $\alpha$ increases (Supporting Information, Table S3) and $\kappa_L$ decreases from S to Te. Whether lone-pair stereoactivity can override this mass-and-softening scaling in $Bi_2O_2X$, and whether the basal-plane symmetry breaking of $Bi_2O_2S$ imposes the same anisotropy on expansion ($\alpha_a \neq \alpha_b$) and on conduction ($\kappa_a \neq \kappa_b$), remain open[40]. To our knowledge, no direction-resolved CTE has been reported for any $Bi_2O_2X$ compound, which precludes estimates of device mismatch strain[38;41]. Thermal transport data, concentrated on $Bi_2O_2Se$, are also inconsistent. First-principles values span 1.3–4.5 $\mathrm{W\,m^{-1}\,K^{-1}}$ in plane and 0.8–3.7 $\mathrm{W\,m^{-1}\,K^{-1}}$ cross plane[37;38;41;42;46;47], and measured values span 1.4–2.4 and 0.9–2.3 $\mathrm{W\,m^{-1}\,K^{-1}}$, respectively[41;48;50–55], because measurements on polycrystalline pellets, supported nanosheets, or along a single axis cannot resolve the full tensor of one crystal across the 300–500 K window relevant to self-heating[8–10]. Validated anisotropic baselines for $\alpha$ and $\kappa_L$ are therefore missing, and the thermomechanical reliability envelope of $Bi_2O_2X$ devices remains undefined[56;57].

Theory has not closed these gaps. For thermal expansion, the quasiharmonic approximation (QHA) retains only volume-dependent harmonic frequencies and omits the intrinsic anharmonic renormalization at fixed volume, whereas the O sublattice of the $[Bi_2O_2]^{2+}$ slabs vibrates above 7 THz ($h\nu/k_B > 330$ K) and therefore retains nuclear quantum effects (NQEs) up to intermediate temperatures[56–58]. For heat transport, early family-wide calculations relied on static 0 K force constants within the relaxation-time approximation or the lowest-order three-phonon (3ph) Boltzmann transport equa-tion (BTE)[38;41;42;47]. Later studies added 4ph scattering for $Bi_2O_2Se$[37;44] or wave-like coherence for $Bi_2O_2S$ at the 3ph level[40], while other studies treated individual compounds or subsets of the family with more restricted transport models[38;41;47;59]. Consequently, even the sign of the $\kappa_L$ difference between $Bi_2O_2S$ and $Bi_2O_2Se$ remains unsettled[40], and no study has combined temperature-renormalized force constants, quartic anharmonicity, and off-diagonal coherent transport across the series. The Wigner transport equation (WTE) provides this framework by adding, to the particle-like population conductivity, a coherence conductivity that becomes significant when phonon linewidths approach interbranch frequency spacings[60]. At the device scale, $Bi_2O_2Se$ FinFETs combine ultrathin fins with high drive currents[11;14], yet no hotspot or thermoelastic analysis has been reported for the sub-15 nm regime in which Fourier diffusion fails[20;21].

Here, we resolve the anisotropic $\alpha$ and $\kappa_L$ tensors of $Bi_2O_2S$, $Bi_2O_2Se$, and $Bi_2O_2Te$ from 100 to 700 K and trace both to Bi lone-pair stereoactivity. Path-integral molecular dynamics (PIMD)[61;62] driven by a machine-learned potential trained on density functional theory (DFT) data captures anharmonicity and NQEs in the lattice parameters, and a mass-substitution control separates the roles of bonding and atomic mass. Temperature-dependent effective potential (TDEP) force constants[63;64] feed iterative solutions of the BTE and WTE that include 4ph scattering and the coherence conductivity $\kappa_{\mathrm{coh}}$[60]. *In situ* X-ray diffraction (XRD) and dual-axis frequency-domain thermoreflectance (FDTR) from 300 to 500 K on $Bi_2O_2Se$ single crystals benchmark the calculations, reproducing the cross-plane CTE (2.82 measured versus $2.64\times10^{-5}$ K$^{-1}$ calculated) and the room-temperature conductivities ($\kappa_{\mathrm{in}} = 1.74\pm0.16$ and $\kappa_{\mathrm{cross}} = 0.92 \pm 0.14$ W m$^{-1}$ K$^{-1}$ measured versus 1.83 and 0.95 calculated) within 6.4%.

The family departs from mass scaling in two ways. First, the volumetric CTE at 300 K falls from $8.14 \times 10^{-5}$ K$^{-1}$ in $Bi_2O_2S$ to 6.72 and $6.61 \times 10^{-5}$ K$^{-1}$ in $Bi_2O_2Se$ and $Bi_2O_2Te$, although the unit-cell volume grows by 13% from S to Te. Substituting the S mass into $Bi_2O_2Se$ changes its expansion by only 0.6%, whereas lone-pair-driven Bi off-centering in $Bi_2O_2S$ creates an antipolar double-well potential and raises the heat-capacity-weighted Grüneisen parameter to 3.55, compared with 3.31 and 3.04 for the selenide and telluride. The same distortion yields $\alpha_b/\alpha_a = 1.63$ but $\kappa_b/\kappa_a = 1.03$, so basal-plane symmetry breaking appears in expansion but not in conduction. Second, $\kappa_L$ peaks in $Bi_2O_2Se$ rather than in the lightest member, because $Bi_2O_2S$ is limited by strong 3ph and 4ph scattering below 3 THz and by low in-plane group velocities, whereas $Bi_2O_2Te$ is limited by the lowest cross-plane group velocity in the family and exhibits the largest conduction anisotropy ($\kappa_{\mathrm{in}}/\kappa_{\mathrm{cross}} = 3.43$). In $Bi_2O_2Se$, 4ph scattering removes 46.8% of the cross-plane population conductivity at 300 K, and $\kappa_{\mathrm{coh}}$ supplies 18.6% of the cross-plane total at 700 K. Finally, non-gray phonon BTE and thermoelastic simulations of 12 nm wide fins with vertically oriented layers, a gate dielectric, and a substrate show that Fourier diffusion underestimates the hotspot temperature and that $Bi_2O_2Se$ develops the lowest hotspot temperature and thermal-mismatch strain under identical heat loads.

# 2 Results and Discussion

## 2.1 Chalcogen-Dependent Bonding and Lone-Pair Stereoactivity

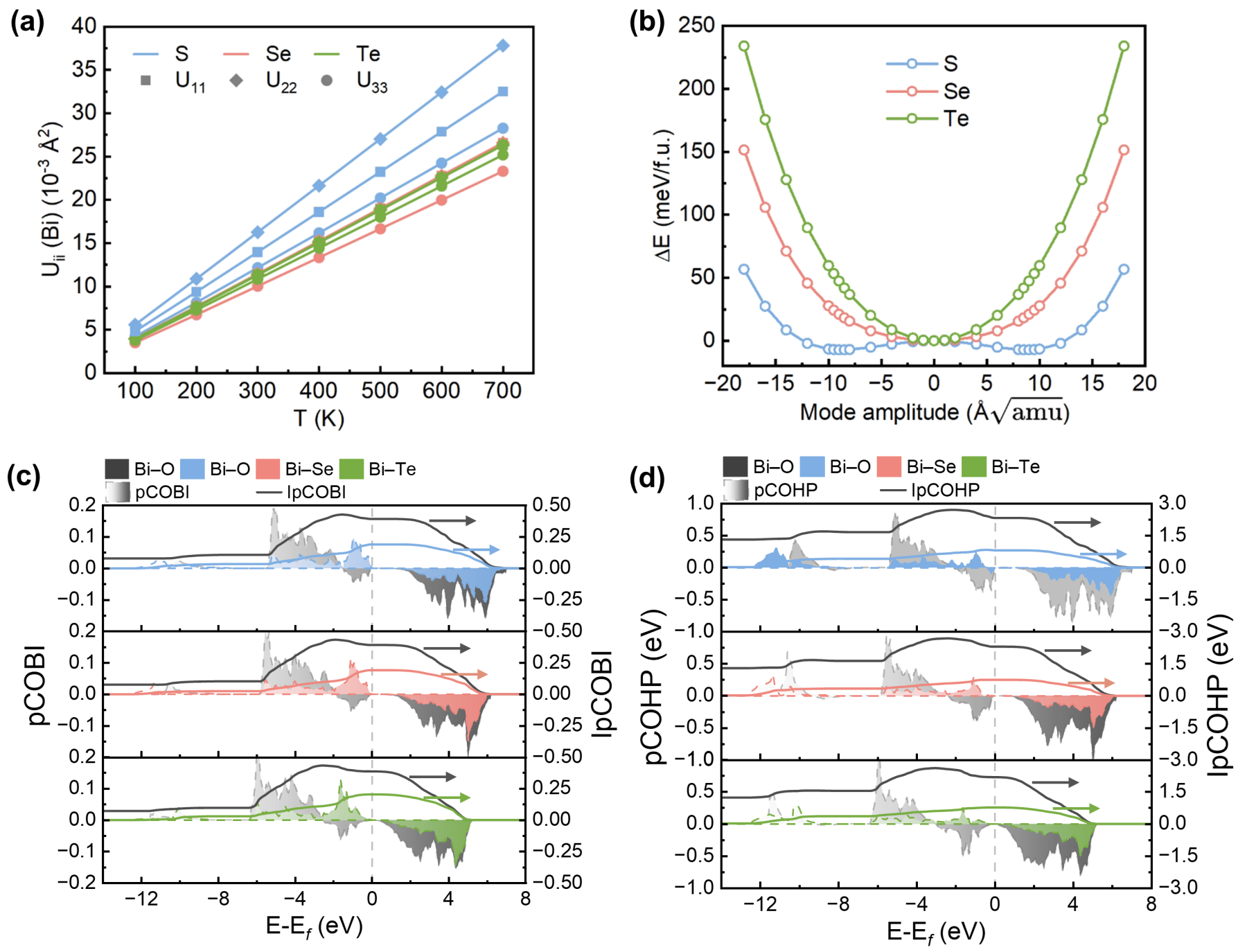

**Figure 2:** Composition-dependent Bi dynamics and chemical bonding. (a) Temperature-dependent diagonal components $U_{ii}$ of the Bi mean-square displacement tensor in $Bi_2O_2S$, $Bi_2O_2Se$, and $Bi_2O_2Te$. (b) Static (0 K) frozen-phonon energy profiles along the antipolar $M$-point mode that connects the tetragonal reference and the orthorhombic structure, evaluated at fixed tetragonal lattice parameters. Energies are referenced to the undistorted configuration and given per formula unit. (c) Bond-averaged pCOBI and IpCOBI and (d) sign-reversed −pCOHP and −IpCOHP for the Bi–O and Bi–X interactions in $Bi_2O_2S$ (top), $Bi_2O_2Se$ (middle), and $Bi_2O_2Te$ (bottom). Gray, blue, red, and green denote Bi–O, Bi–S, Bi–Se, and Bi–Te interactions, respectively. Shaded areas show pCOBI and −pCOHP, and solid curves show their cumulative integrals. Vertical dashed lines mark $E_F$. Integrated values at $E_F$ are listed in Table S1.

The (110)-plane ELF maps in Fig. 1(b) show a pronounced one-sided localization of electron density adjacent to Bi in $Bi_2O_2S$, whereas the corresponding distributions remain nearly symmetric about the centered Bi sites of tetragonal $Bi_2O_2Se$ and $Bi_2O_2Te$ ($I4/mmm$). This asymmetry is the signature of a stereochemically active $Bi^{3+}$ $6s^2$ lone pair and accompanies an in-plane Bi off-centering of $d_{Bi} = 0.454$ Å along $b$ in $Bi_2O_2S$ [Fig. 1(a)], compared with $d_{Bi} = 0$ in the selenide and telluride. The Bi mean-square displacements $U_{ii}$ carry the same symmetry breaking [Fig. 2(a)]. From 100 to 700 K, the displacement amplitudes in $Bi_2O_2S$ exceed those of the other two compounds ($U_{ii}^{S} > U_{ii}^{Te} \gtrsim U_{ii}^{Se}$), indicating a softer local potential around Bi in the sulfide. Fourfold rotational symmetry enforces $U_{11} = U_{22}$ in $Bi_2O_2Se$ and $Bi_2O_2Te$, whereas $Bi_2O_2S$ retains $U_{22} > U_{11} > U_{33}$ throughout the temperature range. At 300 K, $U_{22} = 1.63 \times 10^{-2}$ Å$^2$, compared with $U_{11} = 1.40 \times 10^{-2}$ Å$^2$ and $U_{33} = 1.22 \times 10^{-2}$ Å$^2$, so the largest Bi fluctuations occur along the off-centering direction.

To identify the energetic origin of the distortion, we mapped the static (0 K) energy surface along the symmetry-adapted coordinate $Q$ with frozen-phonon DFT calculations[65] [Fig. 2(b)]. Here, $Q$ is the mass-weighted amplitude of the primitive-cell $M$-point mode that connects the tetragonal $I4/mmm$ reference to the orthorhombic *Pnnm* structure, and all energies were evaluated at fixed, symmetry-constrained tetragonal lattice parameters (Section 4.1.3). Along this coordinate, $Bi_2O_2S$ develops a symmetric double well. A fourth-order fit gives a negative harmonic coefficient, $\Phi^{(2)} = -3.64 \times 10^{-4}$ eV amu$^{-1}$ Å$^{-2}$, minima at $Q = \pm 8.89$ Å amu$^{1/2}$, and a central barrier of 7.21 meV per formula unit. The undistorted structure is therefore unstable at 0 K against antiparallel in-plane displacements of adjacent $[Bi_2O_2]^{2+}$ slabs. Because the dipoles of neighboring slabs cancel, the resulting antipolar *Pnnm* structure remains centrosymmetric, and the positive quartic coefficient $\Phi^{(4)}$ bounds the energy at large $|Q|$. $Bi_2O_2Se$ and $Bi_2O_2Te$ instead exhibit single wells centered at $Q = 0$ with positive $\Phi^{(2)}$ of $3.74 \times 10^{-4}$ and $1.08 \times 10^{-3}$ eV amu$^{-1}$ Å$^{-2}$, respectively, so restoring forces in the heavier chalcogenides prevent condensation of this mode.

The double-well barrier is a static, single-mode quantity. It excludes zero-point and thermal fluctuations, as well as the cell strain and intersite coupling that accompany the cooperative distortion, and therefore cannot be compared directly with $k_B T$ to estimate a transition temperature. The finite-temperature stability of the orthorhombic phase is established instead by PIMD, in which the $a$ and $b$ lattice parameters of $Bi_2O_2S$ remain distinct from 100 to 700 K [Fig. 3(e)], and by the absence of imaginary frequencies in the TDEP dispersions of the *Pnnm* structure at 100 and 700 K [Fig. 5(a)]. This behavior agrees with the orthorhombic ground state reported for $Bi_2O_2S$[45] and with its return to $I4/mmm$ under hydrostatic pressure[40].

Projected crystal orbital Hamilton populations (pCOHP) and crystal orbital bond indices (pCOBI) quantify how the bonding evolves across the series [Fig. 2(c,d) and Table S1]. At the Fermi level, the bond-averaged −ICOHP of Bi–O is 2.32, 2.30, and 2.19 eV per bond in $Bi_2O_2S$, $Bi_2O_2Se$, and $Bi_2O_2Te$, about three times the Bi–X values of 0.81, 0.75, and 0.77 eV per bond. The Bi–O ICOBI remains nearly constant (0.391, 0.391, and 0.385), whereas the Bi–X ICOBI increases from 0.189 and 0.191 for S and Se to 0.203 for Te, indicating a modestly larger covalent contribution to Bi–Te bonding, in line with previous COHP analysis[66]. Because the Bi–X −ICOHP varies non-monotonically, this added covalency does not translate into monotonically stronger Bi–X bonds. The bonding analysis therefore establishes a stiff intralayer Bi–O framework and compliant interlayer Bi–X contacts in all three compounds but does not by itself single out the sulfide.

What distinguishes the sulfide is lone-pair stereoactivity. The deeper S 3$p$ levels lie closer in energy to Bi 6$s$ than the Se 4$p$ or Te 5$p$ levels, strengthening the occupied Bi 6$s$–S 3$p$ antibonding states. When the local symmetry is lowered, these states mix with empty Bi 6$p$ states, lowering the electronic energy and producing the one-sided density in Fig. 1(b)[33;45]. Coupling of this redistribution to antiparallel slab displacements produces the unstable $M$-point mode and the double well in Fig. 2(b). Static off-centering, direction-selective Bi fluctuations, and a shallow anharmonic potential thus form the microscopic basis for the anisotropic expansion and phonon transport analyzed below, paralleling lone-pair-driven anisotropy in strained BiCuSeO[43] and in monolayer $B_4X_4$[67].

## 2.2 Anisotropic Thermal Expansion

### 2.2.1 Phase Verification and *In Situ* X-ray Diffraction

Room-temperature Raman spectroscopy and XRD verified the phase and orientation of the $Bi_2O_2Se$ crystals used in the thermal measurements. The Raman spectrum (Fig. S3) shows the characteristic $A_{1g}$ mode at 159.4 cm$^{-1}$ and the weaker $B_{1g}$ mode near 358 cm$^{-1}$[68;69], consistent with the tetragonal $I4/mmm$ phase.

To benchmark the cross-plane CTE, we performed *in situ* XRD from 298 to 473 K on a crystal supported on a silicon wafer holder [Fig. 3(c)], which suppresses background scattering from the heating stage. Only sharp (002), (004), (006), and (008) reflections appear [Fig. 3(d)], confirming $c$-axis ori-entation, and the resolved Cu $K_{\alpha 1}/K_{\alpha 2}$ doublets indicate high crystalline quality. All (00$l$) reflections shift continuously to lower $2\theta$ on heating, and the room-temperature pattern is recovered after cooling. A linear fit of $c(T)$ between 298 and 398 K, the window that brackets the 300 K PIMD value, gives $\alpha_c = (2.82 \pm 0.12) \times 10^{-5}\,\mathrm{K}^{-1}$, within 6.4% of the PIMD prediction of $2.64 \times 10^{-5}\,\mathrm{K}^{-1}$ (Section 2.2.2). Part of this difference may arise because the fit averages $\alpha_c$ over 298–398 K, whereas PIMD provides the tangent coefficient at 300 K. Residual mechanical coupling between the crystal and the holder may also slightly modify the apparent expansion[70;71].

**Table 1:** Linear and volumetric CTEs ($10^{-5}\,\mathrm{K}^{-1}$) from PIMD and lattice thermal conductivities ($\mathrm{W\,m^{-1}\,K^{-1}}$) at 300 K. Calculated conductivities are WTE values comprising the population term limited by 3ph and 4ph scattering and the coherence term. HNEMD values are listed for $Bi_2O_2S$. The experimental $Bi_2O_2Se$ row gives $\alpha_c$ from *in situ* XRD (linear fit over 298–398 K) and conductivities from dual-axis FDTR. The estimated electronic contribution is below 4% (Section 2.3.3), so the FDTR values represent $\kappa_L$.

| Compound | Space group | $\alpha_a$ | $\alpha_b$ | $\alpha_c$ | $\alpha_V$ | $\alpha$ method | $\kappa_a$ | $\kappa_b$ | $\kappa_{\mathrm{cross}}$ | $\kappa$ method |
|---|---|---|---|---|---|---|---|---|---|---|
| $Bi_2O_2S$ | *Pnnm* | 1.89 | 3.08 | 3.17 | 8.14 | PIMD | 1.04 | 1.07 | 0.50 | WTE |
| $Bi_2O_2S$ | *Pnnm* | — | — | — | — | — | 1.19 | 1.23 | 0.61 | HNEMD |
| $Bi_2O_2Se$ | *I*4/*mmm* | 2.04 | 2.04 | 2.64 | 6.72 | PIMD | 1.83 | 1.83 | 0.95 | WTE |
| $Bi_2O_2Se$, exp. | *I*4/*mmm* | — | — | 2.82 ± 0.12 | — | XRD | 1.74 ± 0.16 | 1.74 ± 0.16 | 0.92 ± 0.14 | FDTR |
| $Bi_2O_2Te$ | *I*4/*mmm* | 1.97 | 1.97 | 2.69 | 6.61 | PIMD | 1.75 | 1.75 | 0.51 | WTE |

#### 2.2.2 Direction-Resolved Expansion from Path-Integral Molecular Dynamics

Isothermal–isobaric PIMD driven by neuroevolution potentials (NEPs) yields the equilibrium lattice parameters $a$, $b$, and $c$ from 100 to 700 K [Fig. 3(e,f)], from which the linear CTEs were obtained [Fig. 3(a) and Table 1]. All lattice parameters increase monotonically with temperature. At 300 K, orthorhombic $Bi_2O_2S$ has $a = 3.89$, $b = 3.96$, and $c = 12.14$ Å ($V \approx 187.1$ Å$^3$), tetragonal $Bi_2O_2Se$ has $a = b = 3.94$ and $c = 12.45$ Å ($V \approx 193.7$ Å$^3$), and tetragonal $Bi_2O_2Te$ has $a = b = 4.03$ and $c = 12.95$ Å ($V \approx 210.8$ Å$^3$), following the chalcogen ionic radius ($r_{\mathrm{S}^{2-}} \approx 1.84$, $r_{\mathrm{Se}^{2-}} \approx 1.98$, and $r_{\mathrm{Te}^{2-}} \approx 2.21$ Å).

The bonding hierarchy of Fig. 2(c,d) accounts for the general anisotropy $\alpha_c > \alpha_{\mathrm{in}}$. The stiff intralayer Bi–O framework restrains basal-plane expansion, whereas the weaker Bi–X contacts provide compliant interlayer connectivity. All linear coefficients are positive, and $\alpha_c/\alpha_{\mathrm{in}}$ remains between 1.28 and 1.37. The electrostatic interlayer contact therefore behaves as a compliant but not membrane-like link, in contrast to graphite, where weakly coupled flexural modes with negative Grüneisen parameters contract the basal plane[31].

The orthorhombic and tetragonal members differ markedly [Table 1]. In $Bi_2O_2S$, the active lone pair breaks basal-plane fourfold symmetry, and the resulting Bi off-centering along $b$ [Figs. 1(a,b) and 2(a)] produces $\alpha_a = 1.89$ and $\alpha_b = 3.08 \times 10^{-5}\,\mathrm{K}^{-1}$ ($\alpha_b/\alpha_a = 1.63$), together with the largest cross-plane coefficient in the family ($\alpha_c = 3.17 \times 10^{-5}\,\mathrm{K}^{-1}$, $\alpha_c/\alpha_a = 1.68$). The most expansive in-plane direction thus coincides with the off-centering direction and the largest Bi fluctuations ($U_{22}$). In tetragonal $Bi_2O_2Se$ and $Bi_2O_2Te$, centered Bi sites preserve $\alpha_a = \alpha_b$, with in-plane coefficients of 2.04 and $1.97 \times 10^{-5}\,\mathrm{K}^{-1}$ and cross-plane coefficients of 2.64 and $2.69 \times 10^{-5}\,\mathrm{K}^{-1}$ ($\alpha_c/\alpha_{\mathrm{in}} = 1.29$ and 1.37), respectively. The larger anisotropy of the telluride is consistent with the heavier Te lowering the frequencies of $c$-polarized acoustic and interlayer breathing modes, which increases their thermal population at 300 K.

### 2.2.3 Inverted Volumetric Expansion and Mass-Substitution Control

At 300 K, PIMD yields $\alpha_V = 8.14$, 6.72, and $6.61 \times 10^{-5}\,\mathrm{K}^{-1}$ for $Bi_2O_2S$, $Bi_2O_2Se$, and $Bi_2O_2Te$, respectively [Fig. 3(b), Fig. S2, and Table S3], so the volumetric expansion decreases as the chalcogen becomes heavier and the cell larger. The decrease is concentrated between S and Se (17%), whereas the selenide and telluride differ by only 1.6%. The anomaly is therefore an enhanced expansion of the sulfide rather than a progressive trend. The chalcogenide series compiled in Fig. 3(b), mainly binary and multinary semiconductors with ionic or mixed ionic–covalent bonding, predominantly show $\alpha_V$ increasing from S to Se and Te, accompanied by lattice softening and weaker cohesion (Table S3b). ZnX[72] and $AgGaX_2$[73] are exceptions, but $Bi_2O_2X$ shows a larger decrease, with $\alpha_V(\mathrm{Se})/\alpha_V(\mathrm{S}) = 0.83$ and $\alpha_V(\mathrm{Te})/\alpha_V(\mathrm{S}) = 0.81$, outside the envelope of the prevailing trend.

To separate atomic mass from bonding, we repeated the PIMD simulations of $Bi_2O_2Se$ with the Se mass (78.97 amu) replaced by the S mass (32.06 amu) while retaining the $Bi_2O_2Se$ potential. In classical statistical mechanics, equilibrium lattice parameters depend only on the potential energy surface, so atomic masses influence thermal expansion solely through nuclear quantum effects. The mass substitution changes $\alpha_V$ by only 0.6% at 300 K relative to $Bi_2O_2Se$ (Table S3a), whereas $Bi_2O_2S$ exceeds $Bi_2O_2Se$ by 21%. The enhanced expansion of $Bi_2O_2S$ therefore resides in its potential energy surface, shaped by lone-pair stereoactivity and the resulting orthorhombic distortion, and not in the lighter chalcogen. Section 2.4.2 relates this result to the heat-capacity-weighted Grüneisen parameters.

Classical molecular dynamics (MD) with the same potentials yields linear and volumetric CTEs 1.2–4.7% larger than PIMD while preserving the anisotropy and compositional trends (Table S3a). This modest reduction reflects the quantum suppression of the heat capacity of the light O atoms in the $[Bi_2O_2]^{2+}$ slabs, whose vibrations lie above 7 THz [Fig. 5(d)], and it bounds the total quantum correction to $\alpha_V$, consistent with the small mass-substitution effect. By contrast, the QHA overestimates $\alpha_V$ by about 17% for all three compounds (9.50, 7.89, and $7.70 \times 10^{-5}\,\mathrm{K}^{-1}$). Because the QHA retains only volume-dependent harmonic frequencies, it omits the explicit anharmonic renormalization at fixed volume that PIMD and MD include, which substantially reduces the expansion in this family.

The cross-plane CTEs of $Bi_2O_2X$ ($2.64$–$3.17 \times 10^{-5}\,\mathrm{K}^{-1}$) exceed that of layered $Bi_2Te_3$ ($\alpha_c \approx 2.2 \times 10^{-5}\,\mathrm{K}^{-1}$)[80] and are an order of magnitude larger than that of silicon ($2.6 \times 10^{-6}\,\mathrm{K}^{-1}$)[74]. Combined with the in-plane inequivalence of $Bi_2O_2S$, this anisotropy concentrates stress at dielectric, contact, and substrate interfaces when $Bi_2O_2X$ is integrated into transistor channels or thermoelectric legs [Fig. 1(c)], which we quantify in Section 2.5.

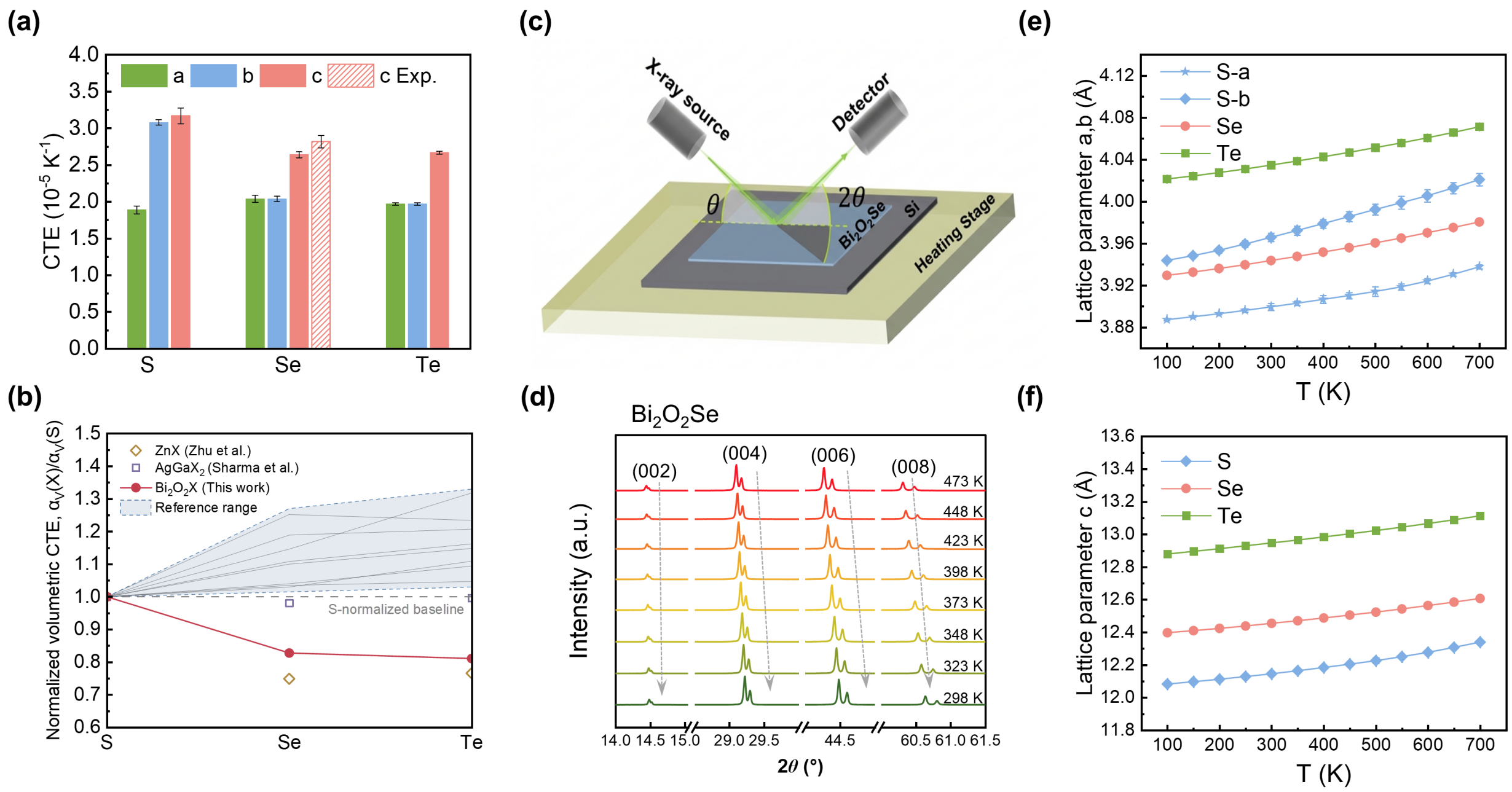

**Figure 3:** Anisotropic and composition-dependent thermal expansion of $Bi_2O_2X$. (a) PIMD linear CTEs along the $a$, $b$, and $c$ axes of $Bi_2O_2S$, $Bi_2O_2Se$, and $Bi_2O_2Te$ at 300 K. The hatched bar denotes the experimental $c$-axis CTE of $Bi_2O_2Se$. (b) Sulfide-normalized volumetric CTEs, $\alpha_V(X)/\alpha_V(\mathrm{S})$, of $Bi_2O_2X$ and representative S–Se–Te chalcogenide series at 300 K. Thin gray lines denote series following the prevailing increase in $\alpha_V$ from S to Se and Te, and the blue shaded region shows their envelope. ZnX[72] and $AgGaX_2$[73], which deviate from this trend, are shown separately. The dashed line marks the sulfide baseline. Values, sources, and trend assignments are given in Table S3b. (c) Schematic of the *in situ* XRD measurement of a $Bi_2O_2Se$ crystal supported on a silicon wafer holder. (d) Temperature-dependent XRD patterns of $Bi_2O_2Se$ from 298 to 473 K showing the shifts of the (002), (004), (006), and (008) reflections. (e,f) PIMD lattice parameters versus temperature, showing (e) $a$ and $b$ and (f) $c$ for the three compounds. Error bars denote 95% confidence intervals from five independent trajectories.

### 2.3 Lattice Thermal Conductivity

#### 2.3.1 Directional and Compositional Trends

Using TDEP force constants with iterative solutions of the WTE, which combine the population conductivity $\kappa_{\mathrm{pop}}$ limited by 3ph and 4ph scattering with the coherence conductivity $\kappa_{\mathrm{coh}}$, we computed $\kappa_{\mathrm{in}}$ and $\kappa_{\mathrm{cross}}$ from 100 to 700 K [Fig. 4(c,d)]. All three compounds have ultralow $\kappa_L$, and in each the cross-plane direction, which expands most, conducts least. The compositional dependence is non-monotonic in both directions, with $\kappa_{\mathrm{in}}(Bi_2O_2Se) > \kappa_{\mathrm{in}}(Bi_2O_2Te) > \kappa_{\mathrm{in}}(Bi_2O_2S)$ and $\kappa_{\mathrm{cross}}(Bi_2O_2Se) > \kappa_{\mathrm{cross}}(Bi_2O_2Te) \approx \kappa_{\mathrm{cross}}(Bi_2O_2S)$ [Fig. 4(a) and Table 1]. This ordering departs from Leibfried–Schlömann–Slack (LSS) scaling[81;82], in which heavier atoms and larger cells lower the Debye temperature and sound velocities and thereby $\kappa_L$, and it differs from the monotonic decrease of $\alpha_V$. Cubic boron pnictides also violate mass scaling[83–85], but there the anomaly arises from mass-ratio-induced restrictions of the scattering phase space. In $Bi_2O_2X$, the low in-plane conductivity of the sulfide instead reflects reduced in-plane group velocities combined with enhanced low-frequency anharmonic scattering associated with the active lone pair (Section 2.4).

Orthorhombic $Bi_2O_2S$ has the lowest in-plane conductivities in the family, $\kappa_a = 1.04$ and $\kappa_b = 1.07\,\mathrm{W\,m^{-1}\,K^{-1}}$ at 300 K, despite S being the lightest chalcogen, and a cross-plane conductivity of $0.50\,\mathrm{W\,m^{-1}\,K^{-1}}$ ($\kappa_a/\kappa_{\mathrm{cross}} = 2.08$ and $\kappa_b/\kappa_{\mathrm{cross}} = 2.14$). It combines the lowest mean in-plane group velocity in the family ($0.466\,\mathrm{km\,s^{-1}}$) with the largest acoustic Grüneisen parameters and the strongest 3ph and 4ph scattering below 3 THz (Sections 2.4.2 and 2.4.3), so reduced velocities and short lifetimes act together.

Although $\alpha_b/\alpha_a = 1.63$ in $Bi_2O_2S$, its in-plane conductivities differ by only 3% ($\kappa_b/\kappa_a = 1.03$). Expansion and conduction weight the lattice dynamics differently. Expansion along $b$ reflects the strain sensitivity of the soft Bi modes polarized along the off-centering direction, whereas conduction integrates $C_V v^2 \tau$ over the entire spectrum. The off-centering raises the mean $b$-axis group velocity by 12% relative to the $a$ axis (0.494 versus $0.438\,\mathrm{km\,s^{-1}}$), and the 3ph-only population conductivity is correspondingly larger along $b$. However, 4ph scattering removes 57.8% of the population conductivity along $b$ but only 29.5% along $a$ at 300 K (Table S4), because the soft Bi modes that carry $b$-polarized heat current have the largest 4ph phase space (Section 2.4.3). The velocity advantage and the additional quartic resistance nearly cancel, so the basal-plane symmetry breaking appears in expansion but not in conduction.

Homogeneous nonequilibrium molecular dynamics (HNEMD) simulations with the same NEP, which include anharmonicity to all orders but treat the dynamics classically, provide an independent check for $Bi_2O_2S$ at 300 and 500 K [Fig. 4(a,c,d) and Table 1]. At 300 K, HNEMD yields $\kappa_a = 1.19$, $\kappa_b = 1.23$, and $\kappa_{\mathrm{cross}} = 0.61\,\mathrm{W\,m^{-1}\,K^{-1}}$, 14–22% above the WTE values. Both methods reproduce nearly equal $a$- and $b$-axis conductivities, a cross-plane conductivity about half of the in-plane values, and a decrease of all three components from 300 to 500 K. The systematic offset is consistent with the classical mode occupations sampled by HNEMD, compared with the quantum statistics of the WTE. Even with this offset, the HNEMD in-plane values of $Bi_2O_2S$ remain well below the WTE value of $Bi_2O_2Se$ ($1.83\,\mathrm{W\,m^{-1}\,K^{-1}}$), so the ordering $\kappa_{\mathrm{in}}(Bi_2O_2S) < \kappa_{\mathrm{in}}(Bi_2O_2Se)$ does not depend on the perturbative treatment of anharmonicity.

Tetragonal $Bi_2O_2Se$ has the highest conductivities in the family along both directions ($\kappa_{\mathrm{in}} = 1.83$ and $\kappa_{\mathrm{cross}} = 0.95\,\mathrm{W\,m^{-1}\,K^{-1}}$ at 300 K, $\kappa_{\mathrm{in}}/\kappa_{\mathrm{cross}} = 1.93$). Its centered Bi sites preserve fourfold symmetry and confine the 4ph phase space relative to the sulfide, and its mean group velocities are the highest in the family ($v_{\mathrm{in}} = 0.615$ and $v_{\mathrm{cross}} = 0.522\,\mathrm{km\,s^{-1}}$). In tetragonal $Bi_2O_2Te$, $\kappa_{\mathrm{in}}$ remains high ($1.75\,\mathrm{W\,m^{-1}\,K^{-1}}$), close to the selenide, whereas $\kappa_{\mathrm{cross}}$ drops to $0.51\,\mathrm{W\,m^{-1}\,K^{-1}}$, matching the sulfide and giving the largest anisotropy in the family ($\kappa_{\mathrm{in}}/\kappa_{\mathrm{cross}} = 3.43$). Weak scattering and an intermediate in-plane velocity ($0.531\,\mathrm{km\,s^{-1}}$) sustain in-plane transport, whereas the heavy Te atoms (127.6 amu) and the compliant interlayer contact reduce $v_{\mathrm{cross}}$ to $0.249\,\mathrm{km\,s^{-1}}$, the lowest in the family, creating a cross-plane velocity bottleneck.

These conductivities ($0.18$–$1.83\,\mathrm{W\,m^{-1}\,K^{-1}}$ over 300–700 K) place $Bi_2O_2X$ among ultralow-conductivity solids such as SnSe and $Bi_2Te_3$ [37–43]. Their anisotropy (1.93–3.43) is moderate compared with that of vdW crystals such as graphite (∼300)[86], $MoS_2$ (∼30)[87], $WSe_2$ (∼10)[88], 2$H$-$MoTe_2$ (7.6)[89], 1$T'$-$MoTe_2$ (7.2), and $T_d$-$MoTe_2$ (5.6)[90;91], reflecting the electrostatic interlayer bonding. It is nonetheless large enough to shape heat removal in nanoscale devices, depending on how the layers are oriented relative to the heat sinks (Section 2.5).

### 2.3.2 Four-Phonon Scattering and Wigner Coherence

Four-phonon scattering and wave-like interband coherence act selectively along the two transport directions (Tables S4 and S5). At 300 K, 4ph scattering reduces $\kappa_{\mathrm{pop}}$ of $Bi_2O_2Se$ from 2.21 to $1.69\,\mathrm{W\,m^{-1}\,K^{-1}}$ in plane (23.6%) and from 1.64 to $0.87\,\mathrm{W\,m^{-1}\,K^{-1}}$ cross plane (46.8%), increasing the population anisotropy from 1.34 to 1.93. This selectivity arises because the cross-plane heat current is concentrated in low-frequency acoustic and optical modes below 3 THz, where large Grüneisen parameters and a large multiphonon phase space make higher-order scattering strongest (Section 2.4.3).

From 100 to 700 K, the in-plane 4ph reduction in $Bi_2O_2Se$ increases from 13.4% to 31.4% as multiphonon channels become thermally accessible, whereas the cross-plane reduction remains larger throughout, decreasing from 49.4% at 100 K to 46.8% at 300 K and 33.1% at 700 K (Table S5). Four-phonon scattering therefore enhances the population anisotropy over the entire temperature range, although its relative cross-plane effect weakens at high temperature as the spectral distribution of the heat current shifts.

The coherence conductivity provides an additional off-diagonal channel[60]. The in-plane $\kappa_{coh}$ of $Bi_2O_2Se$ remains near 0.14 W m$^{-1}$ K$^{-1}$ between 100 and 700 K, whereas the cross-plane value increases by more than a factor of 2.5, from 0.038 at 100 K to 0.074 at 300 K and 0.096 W m$^{-1}$ K$^{-1}$ at 700 K. Thermal broadening of the optical linewidths increasingly bridges the small frequency spacings between adjacent cross-plane optical branches, enhancing their interband overlap.

Because $\kappa_{pop}$ of $Bi_2O_2Se$ decays approximately as $T^{-1.02}$ under combined Umklapp and 4ph scattering (Fig. S6), the coherence fraction $\kappa_{coh}/\kappa_L$ grows with temperature, from 2.5% at 100 K to 7.9% at 300 K and 15.1% at 700 K in plane, and from 1.3% to 7.8% and 18.6% cross plane (Table S5). The coherence channel thus partially offsets the loss of population transport, and the total conductivity follows $T^{-0.95}$ in plane and $T^{-0.92}$ cross plane, consistent with the near-$T^{-1}$ dependence measured here between 300 and 500 K (Section 2.3.3) and reported previously for $Bi_2O_2Se$[48;52]. The apparent near-$T^{-1}$ scaling therefore results from two opposing contributions rather than from a single Umklapp-limited channel. Because $\kappa_{coh}$ increases with linewidth in this regime, strategies that lower $\kappa_L$ by adding scattering, such as alloying or point-defect engineering, will be partly offset by the coherence channel at thermoelectric operating temperatures.

Across the series at 300 K, the cross-plane 4ph reduction decreases from 64.8% in $Bi_2O_2S$ to 46.8% in $Bi_2O_2Se$ and 27.4% in $Bi_2O_2Te$, and the cross-plane $\kappa_{coh}$ decreases from 0.174 to 0.074 and 0.048 W m$^{-1}$ K$^{-1}$. The sulfide thus combines the strongest higher-order resistance with the largest coherence channel, both reflecting the broad linewidths produced by its lone-pair-driven anharmonicity. At 700 K, $\kappa_{coh}$ supplies 78.5% of the cross-plane conductivity of $Bi_2O_2S$ because the population term has nearly vanished. In this regime, the linewidths of many low-frequency modes approach the Ioffe–Regel limit and the perturbative description of the double-well dynamics becomes less reliable, so this fraction should be regarded as semi-quantitative. By contrast, the similarly low $\kappa_{cross}$ of $Bi_2O_2Te$ at 300 K originates mainly from its low cross-plane group velocity, with only moderate 4ph suppression. $Bi_2O_2S$ is therefore predominantly scattering-limited, $Bi_2O_2Te$ is velocity-limited across the layers, and $Bi_2O_2Se$ combines relatively high velocities with moderate higher-order scattering.

### 2.3.3 Dual-Axis Thermoreflectance Benchmark

We measured $\kappa_{in}$ and $\kappa_{cross}$ of cleaved $Bi_2O_2Se$ single crystals (thickness $> 200\,\mu$m) from 300 to 500 K by dual-axis FDTR with a 61.5 nm Pt transducer [Fig. 4(b), inset]. Sweeping the pump modulation frequency $f$ from 0.5 to 10 MHz reduces the thermal penetration depth, $d = (D/\pi f)^{1/2}$, where $D$ is the thermal diffusivity, from about 0.5 to 0.1 $\mu$m cross plane and from about 0.7 to 0.2 $\mu$m in plane. Because these depths remain below the 1.94 $\mu$m beam radius, the response at high frequencies is dominated by cross-plane conduction and the Pt/$Bi_2O_2Se$ interface, whereas radial spreading contributes increasingly at low frequencies. The low thermal conductivity of Pt, relative to the more common Au, limits lateral spreading within the transducer that would otherwise mask the in-plane response of the sample. The resulting sensitivities to $\kappa_{in}$ and $\kappa_{cross}$ have distinct frequency dependences (Fig. S4), allowing both to be fitted simultaneously to a single phase spectrum, a procedure validated on anisotropic reference crystals[92].

At 300 K, the fit yields $\kappa_{in} = 1.74 \pm 0.16$ and $\kappa_{cross} = 0.92 \pm 0.14$ W m$^{-1}$ K$^{-1}$ [Fig. 4(b)], within 5.2% and 3.3% of the WTE values of 1.83 and 0.95 W m$^{-1}$ K$^{-1}$, and the measured anisotropy ratio of 1.89 matches the calculated ratio of 1.93. FDTR measures the total conductivity. Using the reported electrical conductivity of nominally undoped bulk $Bi_2O_2Se$ and a semiconductor Lorenz number, the Wiedemann–Franz law gives an electronic contribution of approximately 0.04 W m$^{-1}$ K$^{-1}$, below 4% of the measured in-plane value between 300 and 500 K (Section 4.2.4). The measured values therefore represent $\kappa_L$. The slight theoretical overestimate lies within the experimental uncertainty and is consistent with point-defect scattering by Se vacancies or Bi antisites[93;94], which the calculations omit. Between 300 and 500 K, both measured components decrease approximately as $T^{-1}$ and follow the calculated temperature dependence within the experimental uncertainty [Fig. 4(c,d)].

Previous reports show substantial axis-dependent scatter in the room-temperature $\kappa_L$ of $Bi_2O_2Se$ [Fig. 4(e,f)]. First-principles predictions span about 1.3–4.5 W m$^{-1}$ K$^{-1}$ in plane and 0.8–3.7 W m$^{-1}$ K$^{-1}$ cross plane[37;38;41;42;46;47], reflecting differences in force constants, scattering orders, BTE solution schemes, and the treatment of coherence. Measured values vary by about a factor of two, from 1.4 to 2.4 W m$^{-1}$ K$^{-1}$ in plane and from 0.9 to 2.3 W m$^{-1}$ K$^{-1}$ cross plane, across sample forms and measurement configurations[48;50–55]. Previous single-crystal and nanostructure studies resolved only one axis[37;95–97] or were limited to sub-ambient temperatures[52], which prevented cross-checks between axes and coverage of the device-relevant window above 400 K. The present measurements determine both components on the same crystal from 300 to 500 K and agree with the WTE in magnitude, anisotropy, and temperature dependence. This agreement suggests that the spread of earlier predictions stems largely from the treatment of force-constant renormalization, 4ph scattering, and coherence, whereas the experimental spread reflects sample form and single-axis methods.

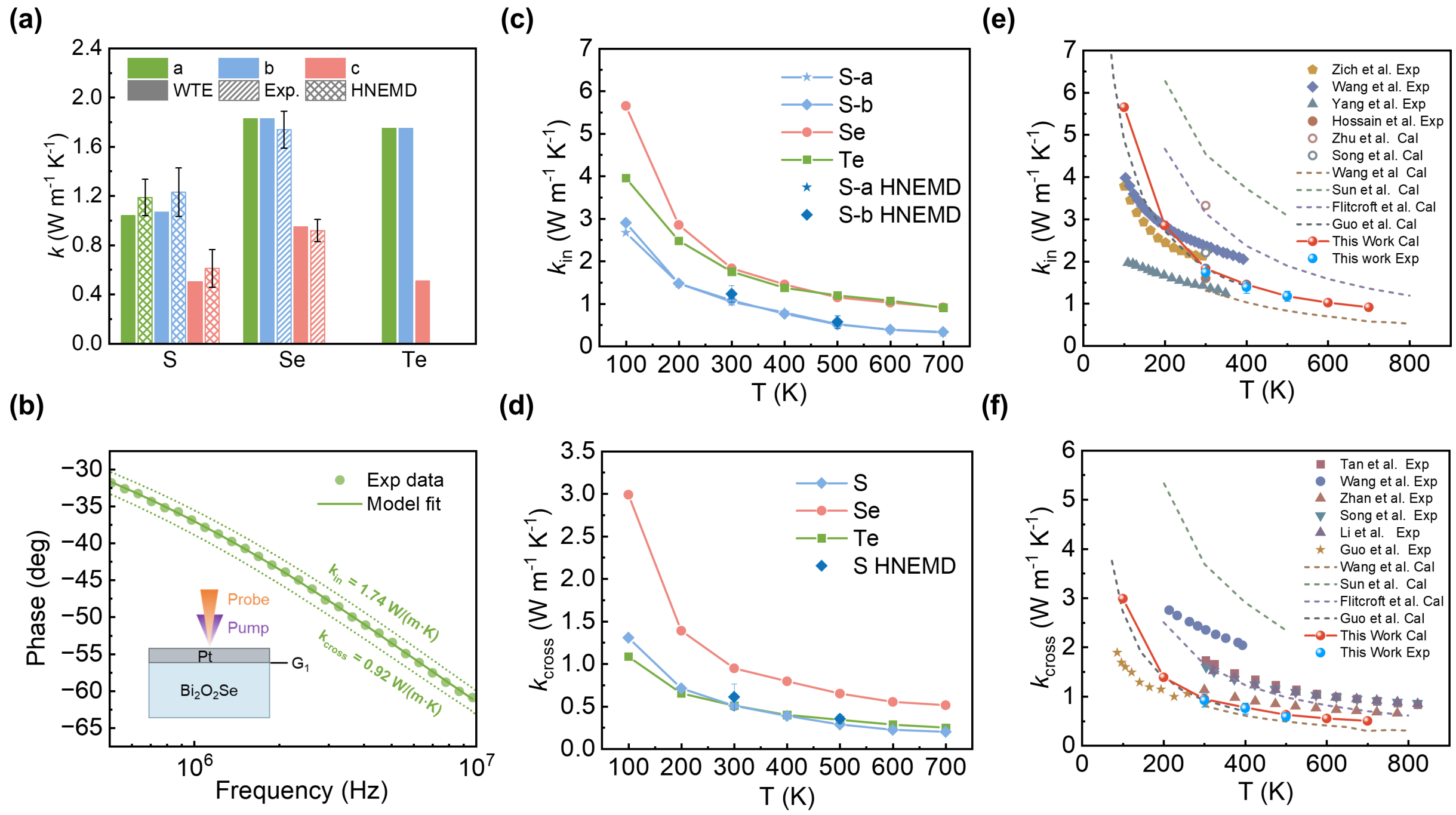

**Figure 4:** Anisotropic thermal conductivity and the $Bi_2O_2Se$ FDTR benchmark. (a) Thermal conductivities along the *a*, *b*, and *c* axes of $Bi_2O_2S$, $Bi_2O_2Se$, and $Bi_2O_2Te$ at 300 K. Solid bars denote WTE results, cross-hatched bars denote HNEMD results for $Bi_2O_2S$, and diagonally hatched bars denote FDTR results for $Bi_2O_2Se$. Error bars indicate the statistical uncertainties of HNEMD and the fitting uncertainties of FDTR. For tetragonal $Bi_2O_2Se$, the *a*- and *b*-axis components are equivalent and correspond to $\kappa_{\text{in}}$. (b) Representative FDTR phase lag of $Bi_2O_2Se$ at 300 K versus modulation frequency. Symbols denote averaged data, and the solid line denotes the simultaneous fit yielding $\kappa_{\text{in}} = 1.74 \pm 0.16$ and $\kappa_{\text{cross}} = 0.92 \pm 0.14\,\text{W}\,\text{m}^{-1}\,\text{K}^{-1}$. The inset shows the pump–probe configuration and the $Pt/Bi_2O_2Se$ thermal model. (c) In-plane and (d) cross-plane thermal conductivities versus temperature, with $Bi_2O_2S$ resolved along *a* and *b* in (c). Lines with symbols denote WTE results, dark-blue symbols with error bars denote HNEMD results for $Bi_2O_2S$ at 300 and 500 K, and open symbols with error bars denote FDTR data for $Bi_2O_2Se$ from 300 to 500 K. (e,f) Comparison of the present (e) in-plane and (f) cross-plane conductivities of $Bi_2O_2Se$ with literature values [37;38;41;42;46–48;52–54] and [37;38;42;47;48;50;51;55], respectively.

## 2.4 Anharmonic Lattice Dynamics and Transport Mechanisms

### 2.4.1 Temperature-Renormalized Phonon Spectra and Group Velocities

The TDEP dispersions at 100 and 700 K [Fig. 5(a–c)] contain no imaginary frequencies, confirming the finite-temperature dynamical stability of all three structures, including the *Pnnm* phase of $Bi_2O_2S$. Because the force constants were sampled on a fixed reference lattice (Section 4.1.2), the renormalization shown here is the explicit anharmonic contribution at constant volume, which hardens most modes from 100 to 700 K, consistent with a dominant positive quartic contribution. Thermal expansion adds a quasiharmonic softening that these force constants exclude. The Raman-active modes of $Bi_2O_2Se$ are reported to soften on heating at ambient pressure [54], indicating that the volume contribution outweighs the explicit shift for those modes. Because the lattice of $Bi_2O_2Se$ expands by about 3% in volume between 300 and 700 K, the calculated conductivities at the highest temperatures should be regarded as upper estimates.

The projected phonon densities of states (PDOS) [Fig. 5(d)] share a two-region structure across the family. Below 6 THz, heavy Bi motion dominates up to about 2.5 THz and chalcogen vibrations dominate between 2 and 5 THz, whereas the manifold above 7 THz consists of O vibrations within the $[Bi_2O_2]^{2+}$ slabs. From $Bi_2O_2S$ to $Bi_2O_2Te$, the maximum phonon frequency decreases from 16.5 to 12.8 THz, and the gap between the two regions widens from about 1.5 to 2.7 THz. Unlike in c-BAs and GaN[83;98], this gap isolates only the O manifold and does not separate the acoustic branches from low-lying optical modes. Below 3 THz, acoustic and optical branches interpenetrate, satisfying the energy- and momentum-conservation conditions for intense multiphonon scattering. For $Bi_2O_2Se$, this low-frequency overlap is consistent with a recent combined DFT, calorimetry, and Raman study that assigned optical modes at 25.7 and 66.5 $cm^{-1}$ (0.77 and 1.99 THz) to shear-type interlayer vibrations and identified strong coupling of the lowest mode to the LA branches along Γ–X and Γ–A[59].

Frequency-resolved group velocities [Fig. 5(e–h)] separate the dispersive acoustic and low-lying optical branches below about 3 THz from the flat, slow O manifold. The mean directional velocities summarize these spectra. For $Bi_2O_2S$, $\upsilon_a = 0.438$ and $\upsilon_b = 0.494\,\mathrm{km\,s^{-1}}$ (in-plane mean $0.466\,\mathrm{km\,s^{-1}}$), and the in-plane means of $Bi_2O_2Se$ and $Bi_2O_2Te$ are 0.615 and 0.531 $\mathrm{km\,s^{-1}}$. This in-plane ordering, $Bi_2O_2Se > Bi_2O_2Te > Bi_2O_2S$, matches that of $\kappa_{\mathrm{in}}$. Because $\kappa_{\mathrm{pop}}$ scales with the squared directional velocity, the low in-plane velocity of $Bi_2O_2S$ reinforces its strong scattering, whereas the weak scattering of $Bi_2O_2Te$ partly compensates for its lower velocity relative to $Bi_2O_2Se$. The cross-plane means follow $Bi_2O_2Se$ ($0.522\,\mathrm{km\,s^{-1}}$) > $Bi_2O_2S$ ($0.298\,\mathrm{km\,s^{-1}}$) > $Bi_2O_2Te$ ($0.249\,\mathrm{km\,s^{-1}}$). The nearly equal $\kappa_{\mathrm{cross}}$ of $Bi_2O_2S$ and $Bi_2O_2Te$ therefore arises from different limitations, namely strong anharmonic scattering in the sulfide and slow cross-plane propagation in the telluride.

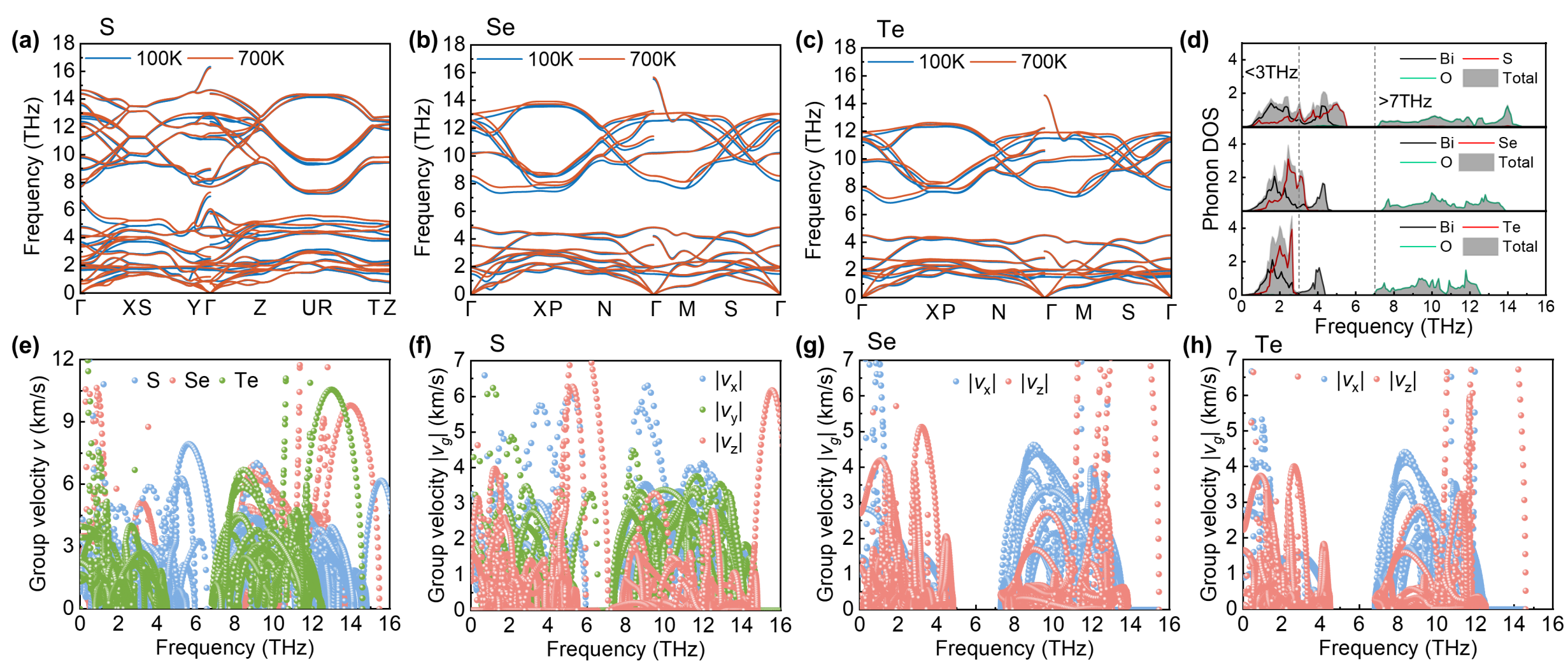

**Figure 5:** Temperature-renormalized phonon dispersions and group velocities of $Bi_2O_2X$. (a–c) TDEP phonon dispersions at 100 and 700 K, obtained at a fixed reference lattice, for (a) $Bi_2O_2S$, (b) $Bi_2O_2Se$, and (c) $Bi_2O_2Te$. (d) Projected phonon densities of states resolving Bi, chalcogen, and O contributions, with dashed lines at 3 and 7 THz. (e) Total phonon group velocity versus frequency. (f–h) Direction-resolved group velocities $|\upsilon_x|$, $|\upsilon_y|$, and $|\upsilon_z|$ for (f) $Bi_2O_2S$, (g) $Bi_2O_2Se$, and (h) $Bi_2O_2Te$.

### 2.4.2 Mode-Resolved Grüneisen Parameters

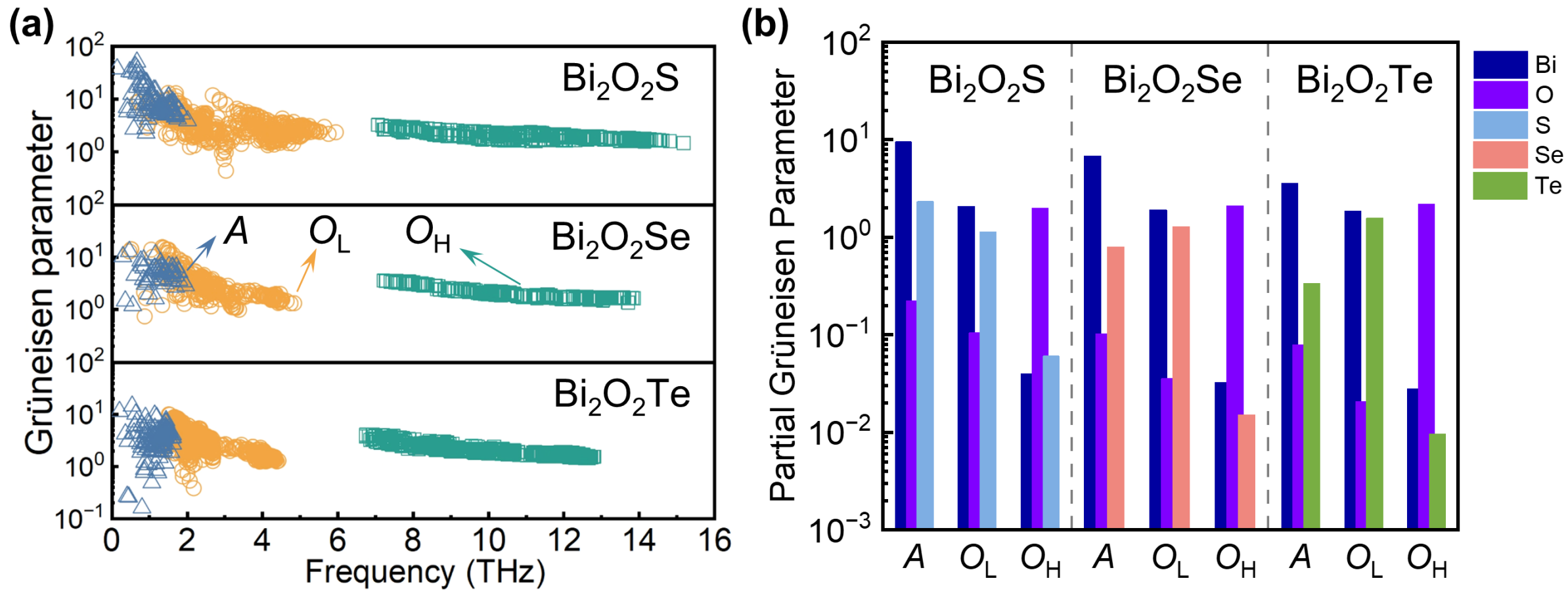

**Figure 6:** Mode-resolved anharmonicity of $Bi_2O_2X$ at 300 K. (a) Volumetric mode Grüneisen parameters versus frequency for $Bi_2O_2S$ (top), $Bi_2O_2Se$ (middle), and $Bi_2O_2Te$ (bottom). (b) Heat-capacity-weighted atomic contributions to the Grüneisen parameters, resolved into acoustic ($A$), low-frequency optical ($O_L$), and high-frequency optical ($O_H$) groups.

The volumetric mode Grüneisen parameters $\gamma_\lambda$ at 300 K (Section 4.1.6) vary most strongly below 3 THz [Fig. 6(a)], where the lowest acoustic and low-lying optical states overlap, whereas the O-dominated manifold above about 6.7 THz shows a narrower distribution of moderate values. The low-frequency response is strongest in $Bi_2O_2S$ and weakens toward $Bi_2O_2Se$ and $Bi_2O_2Te$, consistent with the asymmetric lone-pair density and Bi off-centering of the sulfide [Figs. 1(a,b) and 2]. The active lone pair increases the volume sensitivity of the soft Bi-dominated vibrations, whereas the symmetric Bi environments of the selenide and telluride suppress it.

The heat-capacity-weighted atomic decomposition [Fig. 6(b)] locates the origin of these differences. The $A$-group average decreases from 9.66 in $Bi_2O_2S$ to 7.66 in $Bi_2O_2Se$ and 3.97 in $Bi_2O_2Te$, dominated by Bi, whose partial contribution decreases from 7.52 to 6.77 and 3.56, while the chalcogen contributions are 1.97, 0.79, and 0.33 and the O contributions remain below 0.17. The $O_L$ group varies weakly and non-monotonically (3.28, 3.19, and 3.42), with the Bi contribution decreasing slightly from 2.05 to 1.89 and 1.83 and the chalcogen contribution increasing from 1.12 to 1.26 and 1.57. The $O_H$ group is dominated by O, whose contribution increases slightly from 1.96 to 2.08 and 2.16, with Bi and chalcogen contributions below 0.06. The compositional variation of the low-frequency anharmonicity therefore resides primarily on the Bi sublattice, whereas the high-frequency response belongs to the O sublattice.

These results provide a microscopic interpretation of the PIMD expansion. In the Grüneisen relation $\alpha_V = (B_T V)^{-1} \sum_\lambda C_{V,\lambda}\gamma_\lambda$, where in practice the sum runs over irreducible $\mathbf{q}$ points with symmetry weights $w_\mathbf{q}$, the relevant quantity is the heat-capacity-weighted average $\bar{\gamma}$. At 300 K, $\bar{\gamma}$ decreases from 3.55 in $Bi_2O_2S$ to 3.31 in $Bi_2O_2Se$ and 3.04 in $Bi_2O_2Te$, consistent with $\alpha_V^{\mathrm{S}} > \alpha_V^{\mathrm{Se}} > \alpha_V^{\mathrm{Te}}$. In the classical limit, $\sum_\lambda \gamma_\lambda$ depends only on the force constants, because the mass contribution to the product of phonon frequencies at each $\mathbf{q}$ is independent of volume. The heat-capacity weighting introduces the only mass dependence, consistent with the small effect of mass substitution in PIMD (Section 2.2.3). The $A$ and $O_L$ groups together account for about 80.0%, 77.7%, and 74.5% of $\sum_\lambda C_{V,\lambda}\gamma_\lambda$ in the three compounds, identifying the low-frequency sector as the principal source of expansion, although the $O_H$ contribution remains appreciable at 300 K. The nearly equal $\alpha_V$ of $Bi_2O_2Se$ and $Bi_2O_2Te$ despite their different $\bar{\gamma}$ suggests partial compensation by a smaller bulk modulus in the telluride.

The Grüneisen parameters also gauge the anharmonicity relevant to heat transport, although they do not determine the scattering rates on their own. Within LSS-type models, stronger anharmonicity lowers $\kappa_L$ approximately as $\gamma^{-2}$ [81;82]. The large $A$-group Grüneisen parameter of $Bi_2O_2S$, together with its lowest in-plane velocity, accounts for its lowest in-plane conductivity. $Bi_2O_2Se$ combines a moderate $A$-group response, the smallest $O_L$ average, and the highest velocities along both directions, yielding the largest conductivity. $Bi_2O_2Te$ has the weakest $A$-group anharmonicity but a slightly larger $O_L$ response and lower velocities than $Bi_2O_2Se$, particularly across the layers, so its conductivity remains below that of the selenide. The following section tests these interpretations against mode-resolved scattering rates and phase spaces.

#### 2.4.3 Spectral Transport, Scattering Rates, and Coherence

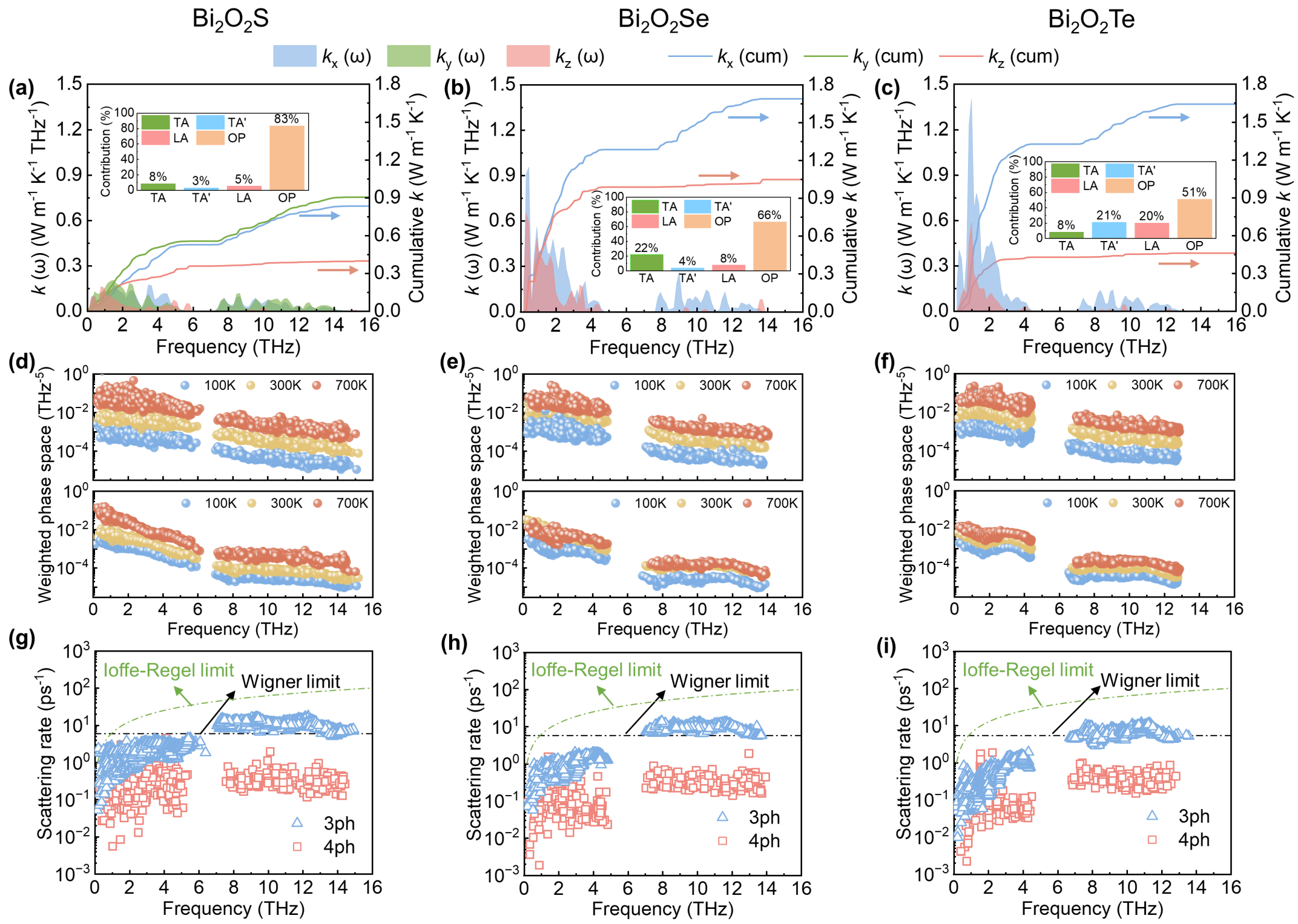

**Figure 7:** Spectral heat transport and anharmonic scattering in $Bi_2O_2X$. (a–c) Spectral thermal conductivity $\kappa(\omega)$ (shaded areas, left axis) and its cumulative integral (solid lines, right axis) along the principal axes of (a) $Bi_2O_2S$, (b) $Bi_2O_2Se$, and (c) $Bi_2O_2Te$. Insets show branch-resolved contributions of the transverse acoustic (TA, TA′), longitudinal acoustic (LA), and optical (OP) branches, defined in the 10-atom cell, which is primitive for *Pnnm* $Bi_2O_2S$ and conventional for *I*4/*mmm* $Bi_2O_2Se$ and $Bi_2O_2Te$. (d–f) Weighted three-phonon (upper panels) and four-phonon (lower panels) scattering phase spaces at 100, 300, and 700 K for (d) $Bi_2O_2S$, (e) $Bi_2O_2Se$, and (f) $Bi_2O_2Te$. (g–i) Mode-resolved 3ph and 4ph scattering rates at 300 K for (g) $Bi_2O_2S$, (h) $Bi_2O_2Se$, and (i) $Bi_2O_2Te$, compared with the Wigner limit (black dashed line) and the Ioffe–Regel limit (green dash-dotted curve).

The spectral accumulation of $\kappa$ [Fig. 7(a–c)] shows how these modes build up the conductivity. In all three compounds, the cumulative conductivity rises steeply below 4 THz, plateaus across the phonon gap between about 5 and 7 THz, and rises again in the O-dominated manifold above 7 THz. The low-frequency rise coincides with the window below 3 THz where acoustic and low-lying optical branches overlap [Fig. 5] and mode Grüneisen parameters peak [Fig. 6(a)]. The branch-resolved contributions in the insets are defined in the 10-atom cell used for all calculations. This cell is primitive for *Pnnm* $Bi_2O_2S$ but conventional for *I*4/*mmm* $Bi_2O_2Se$ and $Bi_2O_2Te$, in which zone folding assigns part of the primitive acoustic dispersion to optical branch indices. Branch fractions are therefore interpreted only within each compound, and comparisons across compounds rely on the frequency-resolved accumulation, which does not depend on the choice of cell. In $Bi_2O_2S$, the three acoustic branches carry only 17% of $\kappa$ (TA, 6%; TA′, 6%; LA, 5%), indicating strongly suppressed acoustic transport.

The weighted phase spaces for three-phonon ($P_3$) and four-phonon ($P_4$) processes [Fig. 7(d–f)] identify the origin of this suppression. Both phase spaces expand with temperature while preserving the two-region structure imposed by the phonon gap. The largest compositional contrast occurs below 3 THz, where $P_4$ of $Bi_2O_2S$ reaches $\sim 10^{-1}$ THz$^{-5}$, exceeding the values of $Bi_2O_2Se$ ($\sim 10^{-2}$ THz$^{-5}$) and $Bi_2O_2Te$ ($\sim 10^{-3}$ THz$^{-5}$) by one and two orders of magnitude, respectively. This enlarged phase space accompanies the soft Bi-centered vibrations identified in Fig. 6(b), the asymmetric lone-pair environment in Fig. 1(b), and the double-well potential in Fig. 2(b), and it is partly enabled by the broken basal-plane symmetry. Consistently, the mode-resolved rates at 300 K [Fig. 7(g–i)] show that below 3 THz $Bi_2O_2S$ has the highest 3ph and 4ph rates, $Bi_2O_2Te$ the lowest, and $Bi_2O_2Se$ an intermediate response. Although 4ph rates are smaller than 3ph rates for most high-frequency modes, 4ph processes act most strongly on the long-lived modes below 3 THz and therefore supply the additional resistance that suppresses cross-plane population transport (Table S4).

Comparing the scattering rates with the Wigner and Ioffe–Regel limits distinguishes the two transport channels [Fig. 7(g–i)]. Below about 5 THz, scattering rates remain well below the Wigner limit ($\Gamma_\lambda \ll \Delta\omega$), so transport is dominated by the population channel of propagating wavepackets. In the dense optical manifold above 7 THz, 3ph rates approach or exceed the interbranch spacings ($\Gamma \sim |\omega_s - \omega_{s'}|$) while remaining below the Ioffe–Regel limit ($\Gamma_\lambda < \omega_\lambda$), the regime in which the off-diagonal coherence term becomes significant[60]. The O-dominated manifold of the $[Bi_2O_2]^{2+}$ slabs is therefore the main source of $\kappa_{coh}$. At 300 K, however, $\kappa_{coh}$ accounts for less than 8% of the conductivity of $Bi_2O_2Se$, so most of the optical contribution at room temperature is carried by the population channel. The coherence share grows as the linewidths broaden with temperature (Section 2.3.2).

Together, these results trace both family trends to the Bi lone pair. In $Bi_2O_2S$, lone-pair activation drives static Bi off-centering and a double-well potential, which raise the heat-capacity-weighted Grüneisen parameter and the volumetric expansion and open a large low-frequency multiphonon phase space. Combined with the lowest in-plane velocity, this makes the lightest compound the poorest in-plane conductor. In $Bi_2O_2Te$, centered Bi sites and weak scattering sustain a high $\kappa_{in}$, whereas the lowest cross-plane velocity limits $\kappa_{cross}$. $Bi_2O_2Se$ combines moderate higher-order scattering with the highest velocities along both directions and conducts best. In all three compounds, the dense optical manifold of the $[Bi_2O_2]^{2+}$ slabs supplies a coherence channel whose share grows with temperature.

## 2.5 Device-Scale Self-Heating and Thermoelastic Response

To translate these material-level differences to device operation, we simulated steady-state phonon transport and the resulting thermoelastic response of a FinFET channel with a fin width of 12 nm, a height of 60 nm, and a length of 50 nm [Fig. 8(a)]. Following fabricated $Bi_2O_2Se$ FinFETs, in which the $[Bi_2O_2]^{2+}$ layers stand vertically[11;14], the *c* axis was aligned with the fin width, so that the channel and fin-height directions both lie in the *ab* plane. The fin is covered on its top and sidewalls by a gate dielectric and rests on a substrate, and the source and drain contacts act as 300 K thermalizing boundaries (Section 4.1.8). Uniform volumetric heat-generation rates of $Q_1 = 1.4\times10^{18}$ and $Q_2 = 5.8\times10^{18}$ W m$^{-3}$ were applied in the drain-side upper region of the fin, corresponding to total powers of approximately 3.94 and 16.34 $\mu$W per fin, or about 0.030 and 0.124 mW $\mu$m$^{-1}$ normalized by the effective gate width $2H + W = 132$ nm. $Q_1$ and $Q_2$ represent the normal and elevated loads, respectively. All results in Fig. 8 were obtained with modal phonon properties evaluated at 300 K, including isotope, 3ph, and 4ph scattering.

### 2.5.1 Composition-Dependent Self-Heating

Under both loads, the peak hotspot temperatures follow $T_{hot}(Bi_2O_2S) > T_{hot}(Bi_2O_2Te) > T_{hot}(Bi_2O_2Se)$ [Fig. 8(b)]. At the elevated load, $T_{hot}$ reaches 659.43 K in $Bi_2O_2S$, 516.14 K in $Bi_2O_2Se$, and 567.09 K in $Bi_2O_2Te$, so $Bi_2O_2Se$ reduces the temperature rise above 300 K by 39.9% relative to $Bi_2O_2S$ and by 19.1% relative to $Bi_2O_2Te$. With vertically oriented layers, heat leaves the hotspot toward the contacts along the channel and toward the substrate along the fin height, both within the *ab* plane, so the ordering follows $\kappa_{in}$. $Bi_2O_2S$, whose in-plane conductivity is about 40% lower than that of $Bi_2O_2Se$, accumulates the most heat. $Bi_2O_2Te$ nearly matches the in-plane conductivity of $Bi_2O_2Se$, whereas its low $\kappa_{cross}$ limits lateral heat redistribution across the 12 nm fin width. The temperature field of $Bi_2O_2Se$ under the normal load [Fig. 8(c)] peaks in the drain-side upper fin and decays toward the contacts and the substrate, and the corresponding heat-flux distribution is shown in Fig. S7.

### 2.5.2 Thermal-Mismatch Strain

The temperature fields were transferred to a thermoelastic model that additionally included the gate dielectric and used the temperature-dependent PIMD CTE tensors (Section 4.1.8). We report the thermal-mismatch strain $\varepsilon_{mis}$, defined as the elastic part of the total strain, which vanishes for a uniformly heated, unconstrained solid. It arises from the incompatibility of nonuniform expansion within the fin and from the expansion mismatch between the channel and the dielectric and substrate. $\varepsilon_{mis}$ is largest in the heated upper fin, where the temperature rise and the mismatch with the dielectric coincide [Fig. 8(d)]. Under the elevated load, its maximum, $\varepsilon_{mis}^{max}$, reaches 4.14%, 1.82%, and 2.48% in $Bi_2O_2S$, $Bi_2O_2Se$, and $Bi_2O_2Te$, respectively, following the same ordering as $T_{hot}$.

The vertical-layer orientation determines which expansion coefficients control this strain. At the sidewall interfaces, whose normal is the *c* axis, the tangential mismatch is set by the in-plane coefficients along the channel and the fin height, whereas $\alpha_c$ acts across the 12 nm width and along the top interface. $Bi_2O_2S$ combines the largest temperature rise with $\alpha_b = 3.08 \times 10^{-5}\ \mathrm{K}^{-1}$ along the fin height, 1.63 times $\alpha_a$ along the channel, producing the strongest and most anisotropic sidewall mismatch. $Bi_2O_2Se$ and $Bi_2O_2Te$ have nearly equal in-plane coefficients ($2.04$ and $1.97 \times 10^{-5}\ \mathrm{K}^{-1}$), so the lower strain of $Bi_2O_2Se$ originates mainly from its smaller temperature rise rather than from a smaller expansion coefficient.

### 2.5.3 Non-Fourier Transport and Model Sensitivity

To isolate non-Fourier effects, the $Bi_2O_2Se$ device under the normal load was also simulated with anisotropic Fourier conduction [Fig. 8(e,f)]. The Fourier model used the bulk conductivity tensor computed from the same modal phonon properties, together with identical thermal geometry, mesh, material properties, heat source, and boundary conditions, so that both models share the same bulk limit (Section 4.1.8). Both temperature fields were subsequently applied to the same thermoelastic model including the gate dielectric. The Fourier model predicts $T_{hot} = 341.37$ K and $\varepsilon_{mis}^{max} = 0.45\%$, compared with 346.2 K and 0.66% from the BTE-based calculation. Two sub-continuum effects account for the higher BTE temperature. First, phonons with mean free paths comparable to or longer than the 12 nm fin width scatter diffusely at the channel–dielectric sidewalls, which lowers the effective in-plane conductivity of the fin along both heat-removal paths below its bulk value. Second, near the thermalizing source and drain boundaries, the phonon distribution departs from local equilibrium, producing temperature jumps that act as additional ballistic resistances absent from the Fourier description. Fourier-based estimates therefore underestimate both the self-heating and the resulting mismatch strain in sub-15 nm $Bi_2O_2X$ fins[20–22].

The BTE solver treats particle-like transport within the relaxation-time approximation and uses a fixed set of modal properties that is not updated with the local temperature, which limits its quantitative accuracy at large temperature rises[21;99;100]. To assess this approximation, the elevated-load calculations were repeated with property sets evaluated at 500 and 700 K (Table S7). Higher-temperature property sets raise $T_{hot}$ for all three compounds, so the absolute elevated-load temperatures depend on the property-evaluation temperature. The ordering $T_{hot}(Bi_2O_2S) > T_{hot}(Bi_2O_2Te) > T_{hot}(Bi_2O_2Se)$, however, is preserved for all three sets. The BTE also omits the coherence channel. Because the coherence contribution is largest in $Bi_2O_2S$ (Section 2.3.2), this omission affects the sulfide most. Even the full WTE in-plane conductivity of $Bi_2O_2S$, which includes coherence, remains about 42% below that of $Bi_2O_2Se$, so the compositional ordering does not depend on this approximation.

Overall, the device simulations translate the material-level trends into a consistent thermomechanical response. With vertically oriented layers, both heat removal and sidewall mismatch are governed mainly by in-plane properties, whereas the low cross-plane conductivity, which would dominate in horizontally stacked channels, plays a secondary role. Under identical geometry and loads, the high $\kappa_{in}$ and isotropic in-plane expansion of $Bi_2O_2Se$ therefore yield the lowest hotspot temperature and mismatch strain among the three compounds. These thermoelastic calculations compare temperature-induced deformation across compositions and are not intended to predict fracture, fatigue, or device lifetime.

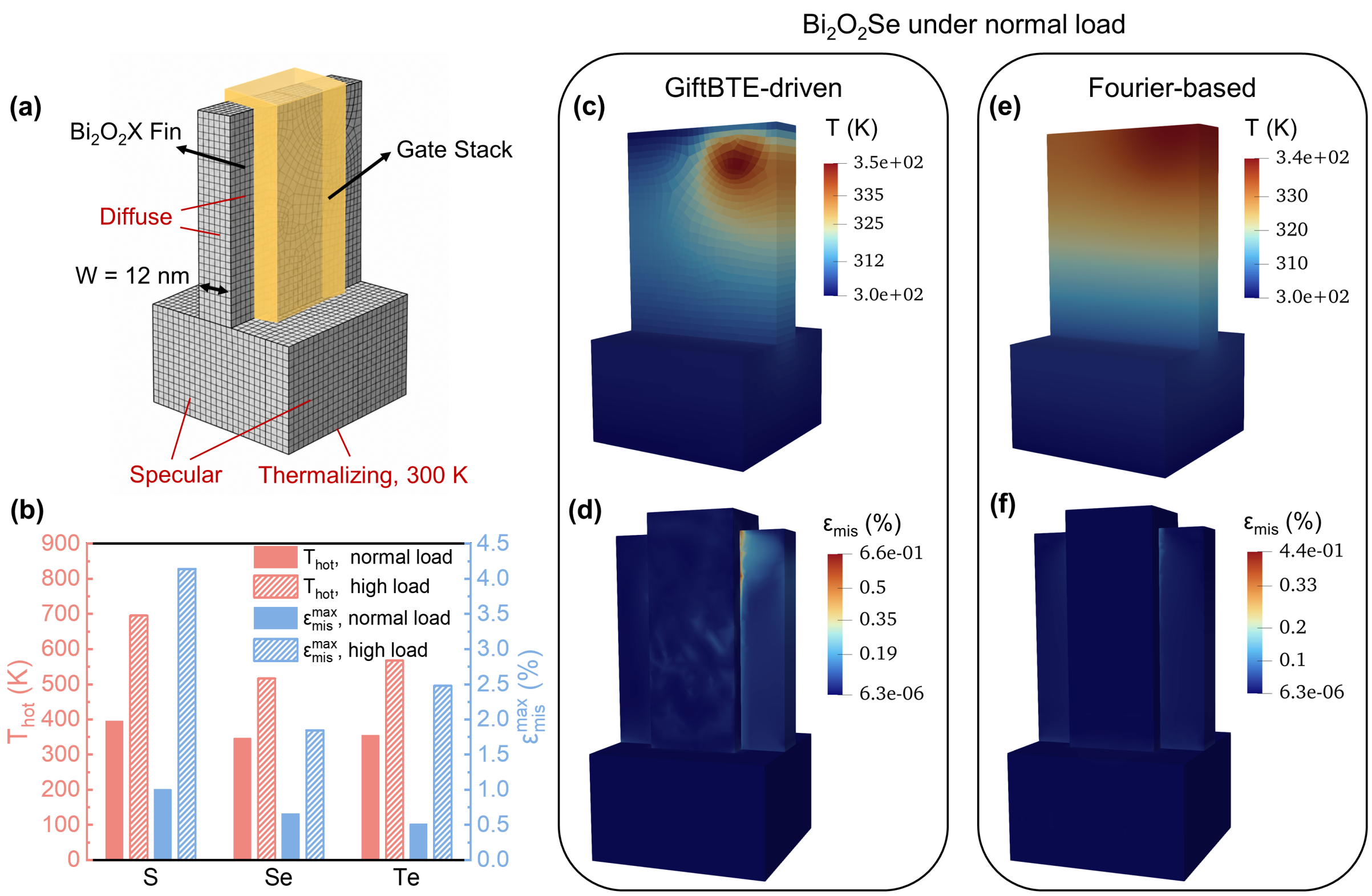

**Figure 8:** Device-scale phonon transport and thermoelastic response of $Bi_2O_2X$ FinFET channels. (a) Device geometry, finite-element mesh, and boundary conditions for the 12 nm wide, 60 nm high, and 50 nm long fin. The $[Bi_2O_2]^{2+}$ layers stand vertically, with the crystallographic $c$ axis along the fin width. The fin rests on a substrate. The gate dielectric covers its top and sidewalls and is included explicitly only in the thermoelastic calculations. Thermalizing boundaries at 300 K are applied at the source and drain faces and at the bottom of the substrate. (b) Peak hotspot temperatures, $T_{hot}$, from the phonon BTE and maximum thermal-mismatch strains, $\varepsilon_{mis}^{max}$, of $Bi_2O_2S$, $Bi_2O_2Se$, and $Bi_2O_2Te$ under the normal load ($Q_1 = 1.4 \times 10^{18}$ W m$^{-3}$) and the elevated load ($Q_2 = 5.8 \times 10^{18}$ W m$^{-3}$), calculated with phonon properties evaluated at 300 K. (c) Steady-state BTE temperature field and (d) corresponding mismatch-strain field of $Bi_2O_2Se$ under the normal load. (e) Fourier temperature field and (f) corresponding mismatch-strain field under the same loading, using the same thermal model geometry as (c) and the same thermoelastic model geometry as (d).

# 3 Conclusions

We have combined path-integral molecular dynamics, Wigner transport calculations based on temperature-dependent force constants with three- and four-phonon scattering, *in situ* X-ray diffraction, and dual-axis frequency-domain thermoreflectance to resolve the anisotropic thermal expansion and lattice thermal conductivity of $Bi_2O_2X$ ($X = S, Se, Te$) and to trace both to the stereochemical activity of the $Bi^{3+}$ $6s^2$ lone pair. In $Bi_2O_2S$, the strong Bi $6s$–S $3p$ antibonding interaction activates the lone pair, drives in-plane Bi off-centering along $b$, and produces an antipolar double-well potential, whereas centered Bi sites preserve tetragonal symmetry in $Bi_2O_2Se$ and $Bi_2O_2Te$.

This lone-pair contrast overrides the mass and softening scaling followed by most chalcogenide series. The volumetric thermal expansion coefficient at 300 K decreases from $8.14 \times 10^{-5}\,K^{-1}$ in $Bi_2O_2S$ to 6.72 and $6.61 \times 10^{-5}\,K^{-1}$ in $Bi_2O_2Se$ and $Bi_2O_2Te$, although the unit cell expands by 13% from S to Te. Substituting the S mass into $Bi_2O_2Se$ changes its expansion by only 0.6%, so the enhanced expansion of the sulfide originates from its potential energy surface, consistent with its larger heat-capacity-weighted Grüneisen parameter (3.55, compared with 3.31 and 3.04). Cross-plane expansion exceeds in-plane expansion throughout the family ($\alpha_c = 2.64$–$3.17 \times 10^{-5}\,K^{-1}$). In $Bi_2O_2S$, the $b$-axis coefficient is 1.63 times the $a$-axis value, whereas the in-plane conductivities differ by only 3% because the higher $b$-axis group velocity is offset by stronger four-phonon scattering. The lattice thermal conductivity peaks in $Bi_2O_2Se$ (1.83 and $0.95\,W\,m^{-1}\,K^{-1}$ in plane and cross plane at 300 K). $Bi_2O_2S$ is limited by strong anharmonic scattering below 3 THz and low in-plane group velocities, whereas $Bi_2O_2Te$ is limited by the lowest cross-plane group velocity in the family and shows the largest conduction anisotropy (3.43). In $Bi_2O_2Se$, four-phonon scattering removes 46.8% of the cross-plane population conductivity at 300 K, and the coherence channel supplies 18.6% of the cross-plane conductivity at 700 K, partly offsetting the decay of population transport and producing the observed near-$T^{-1}$ temperature dependence. X-ray diffraction ($\alpha_c = 2.82 \times 10^{-5}\,K^{-1}$) and thermoreflectance from 300 to 500 K ($\kappa_{in} = 1.74 \pm 0.16$ and $\kappa_{cross} = 0.92 \pm 0.14\,W\,m^{-1}\,K^{-1}$ at 300 K) on $Bi_2O_2Se$ single crystals agree with these predictions within 6.4%.

At the device level, phonon Boltzmann transport simulations of 12 nm wide FinFET channels with vertically oriented layers, a gate dielectric, and a substrate predict higher hotspot temperatures than Fourier diffusion (346.2 K versus 341.37 K for $Bi_2O_2Se$ under the normal load), showing that sub-continuum transport must be included in self-heating estimates at these dimensions. Because heat removal and sidewall mismatch in this orientation are governed mainly by in-plane properties, $Bi_2O_2Se$ develops the lowest hotspot temperature and thermal-mismatch strain under identical loads, followed by $Bi_2O_2Te$ and $Bi_2O_2S$. These results identify lone-pair stereoactivity as a chemical control parameter for both thermal expansion and heat conduction in non-van der Waals layered semiconductors. They also provide validated anisotropic property sets for the thermal and thermomechanical design of $Bi_2O_2X$ transistors, photodetectors, and thermoelectric devices.

# 4 Methods

## 4.1 Computational Methods

### 4.1.1 First-Principles Calculations

Density functional theory (DFT) calculations were performed with the Vienna *ab initio* simulation package (VASP)[101] using the Perdew–Burke–Ernzerhof (PBE) exchange-correlation functional[102]. Structural optimizations used a plane-wave cutoff of 520 eV and convergence thresholds of $10^{-8}$ eV for the total energy and $10^{-6}\,eV\,Å^{-1}$ for the atomic forces. Reference configurations for potential training were generated by *ab initio* molecular dynamics (AIMD) between 50 and 750 K in $4 \times 4 \times 1$ supercells with Γ-point sampling. To improve the transferability of the potentials, uniform triaxial strains from −1.5% to +1.5% and uniaxial strains from −3% to +3% were applied to the equilibrium structures. The energies, forces, and virials of the selected configurations were then recalculated with the same cutoff and a Γ-centered $2 \times 2 \times 3$ $k$-point mesh.

### 4.1.2 Neuroevolution Potentials and Temperature-Dependent Force Constants

Composition-specific NEP models[103;104] were trained for $Bi_2O_2S$, $Bi_2O_2Se$, and $Bi_2O_2Te$ with the GPUMD package[105]. These potentials provide energies, forces, and virials with near-DFT accuracy at a small fraction of the computational cost[2;42;50;55;107]. Each NEP expresses the site energy $U_i$ of atom $i$ as the output of a single-hidden-layer feed-forward neural network whose input is a rotation- and translation-invariant descriptor vector $\mathbf{q}^i = \{q^i_\nu\}_{\nu=1}^{N_{\mathrm{des}}}$ constructed from Chebyshev-based radial and angular components:

$$U_i = \sum_{\mu=1}^{N_{\mathrm{neu}}} w^{(1)}_{\mu} \tanh\left(\sum_{\nu=1}^{N_{\mathrm{des}}} w^{(0)}_{\mu\nu} q^i_\nu - b^{(0)}_{\mu}\right) - b^{(1)}, \tag{1}$$

where $N_{\mathrm{neu}}$ and $N_{\mathrm{des}}$ are the numbers of hidden-layer neurons and descriptor components, and the weights $w$ and biases $b$ were optimized with the separable natural evolution strategy. Atomic forces and virials follow from derivatives of the total energy[105].

For each compound, the training database was assembled from 119 AIMD trajectories, comprising temperature-ramped trajectories from 50 to 750 K and isothermal trajectories from 100 to 700 K in 100 K steps. The final dataset of each compound contained 884 configurations (224 uniaxially strained and 660 triaxially strained) and was randomly divided at the configuration level into training (90%) and test (10%) sets. Each NEP used radial and angular cutoffs of 7 and 5 Å, $n^R_{\max} = 12$, $n^A_{\max} = 8$, 50 hidden neurons, a batch size of 1000, and $2 \times 10^5$ generations. The root-mean-square errors on the training and test sets are $4.5 \times 10^{-4}$ and $3.9 \times 10^{-4}$ eV atom$^{-1}$ for energies, $5.3 \times 10^{-2}$ and $5.2 \times 10^{-2}$ eV Å$^{-1}$ for forces, and $7.1 \times 10^{-3}$ and $7.3 \times 10^{-3}$ eV atom$^{-1}$ for virials, respectively (Fig. S1).

Temperature-dependent second-, third-, and fourth-order interatomic force constants (IFCs) were extracted within the TDEP framework[63;64] using hiPhive[108] from configurations sampled by classical NEP-driven molecular dynamics (MD). At each temperature, a 4×4×2 supercell of the 10-atom cell (320 atoms), fixed at the DFT-relaxed lattice parameters, was equilibrated for 10 ps in the canonical ($NVT$) ensemble with a 1 fs time step, followed by a 50 ps production run from which 100 configurations were sampled at 0.5 ps intervals. The IFCs of all three orders were fitted jointly within a single cluster space with cutoffs of 7.5, 6.5, and 4.5 Å, respectively, by standardized ordinary least squares without regularization, subject to space-group and permutation symmetries and the acoustic sum rules. Because the same reference lattice was used at all temperatures, the TDEP force constants contain the explicit anharmonic renormalization at constant volume but not the effect of thermal expansion. PIMD was used only for equilibrium structural and thermodynamic properties (Section 4.1.7) and not for the transport force constants.

### 4.1.3 Electron Localization, Bi Displacements, Frozen-Phonon Energetics, and Bonding Analysis

The electron localization function (ELF) was calculated from the converged Kohn–Sham wavefunctions of the fully relaxed ground-state structures with the DFT settings described above. Two-dimensional slices on the (110) plane are plotted on a common scale from 0 to 1 for all three compounds. The in-plane Bi off-centering $d_{\mathrm{Bi}}$ was defined as the magnitude of the $ab$-plane projection of the displacement between the Bi position in the relaxed structure and the corresponding site of the symmetry-matched $I4/mmm$ reference. The Bi mean-square displacement tensor $\mathbf{U}(T)$ was calculated from the second-order IFCs with Phonopy[65] from 100 to 700 K in 50 K steps. Its diagonal elements, $U_{ii} = \langle u_i^2 \rangle$ with $i = 1$, 2, and 3 corresponding to the $x$, $y$, and $z$ directions, were averaged over the four Bi atoms of the 10-atom cell.

The energy landscape of the $Pnnm$-forming distortion was evaluated with static frozen-phonon calculations at 0 K. Symmetry-constrained $I4/mmm$ reference structures were first optimized. The distortion coordinate was identified by projecting the displacement between the $I4/mmm$ and $Pnnm$ structures of $Bi_2O_2S$ onto the phonon eigenvectors of the reference, and the primitive-cell $M$-point mode with the largest projection was selected. The corresponding modes of $Bi_2O_2Se$ and $Bi_2O_2Te$ were identified from eigenvector-subspace overlaps. This doubly degenerate $M$-point mode folds to $\Gamma$ in the 10-atom conventional cell, in which its eigenvector was normalized as $\sum_{\kappa\alpha} |e_{\kappa\alpha}|^2 = 1$. The mass-weighted mode coordinate $Q$ was defined through

$$u_{\kappa\alpha}(Q) = \frac{Q\, e_{\kappa\alpha}}{\sqrt{M_\kappa}}, \tag{2}$$

where $u_{\kappa\alpha}$, $e_{\kappa\alpha}$, and $M_\kappa$ are the displacement, eigenvector component, and mass of atom $\kappa$ along direction $\alpha$. Total energies of the displaced structures were computed without ionic or cell relaxation, at the fixed lattice parameters of the reference, using the PBE functional, the Bi_d projector augmented-wave potential, a cutoff of 550 eV, and a $\Gamma$-centered $13 \times 13 \times 4$ $k$-point mesh. The energies were referenced to $Q = 0$, divided by two to obtain values per formula unit, symmetrized between $+Q$ and $-Q$, and fitted for $|Q| \leq 14$ Å amu$^{1/2}$ to

$$\Delta E_{\text{f.u.}}(Q) = \frac{1}{2!}\Phi^{(2)}Q^2 + \frac{1}{4!}\Phi^{(4)}Q^4, \tag{3}$$

where $\Phi^{(2)} = (\partial^2 \Delta E_{\text{f.u.}}/\partial Q^2)_{Q=0}$ is the harmonic curvature at the high-symmetry configuration and $\Phi^{(4)} = (\partial^4 \Delta E_{\text{f.u.}}/\partial Q^4)_{Q=0}$ is the leading anharmonic coefficient along the selected coordinate. For $\Phi^{(2)} < 0$, the minima lie at $Q_0 = \pm(-6\Phi^{(2)}/\Phi^{(4)})^{1/2}$, and the barrier height is $\Delta E_{\text{b}} = 3[\Phi^{(2)}]^2/(2\Phi^{(4)})$. These quantities describe the static energy surface along a single mode and exclude zero-point and thermal fluctuations, cell relaxation, and coupling to other modes.

Chemical bonding was analyzed with LOBSTER[109–113] from static PBE wavefunctions of the relaxed ground-state structures, using the pbeVaspFit2015 basis set with Bi 5$d$, 6$s$, and 6$p$, O 2$s$ and 2$p$, and chalcogen $ns$ and $np$ orbitals. Projected crystal orbital Hamilton populations (pCOHP) and projected crystal orbital bond indices (pCOBI)[114] were calculated from −20 to 10 eV relative to $E_F$, and their cumulative energy integrals are denoted IpCOHP($E$) and IpCOBI($E$). For each compound, the results were averaged over the 16 nearest-neighbor Bi–O and the 16 Bi–X contacts of the 10-atom cell. pCOHP and IpCOHP are plotted with reversed sign, so that positive values denote bonding contributions, whereas pCOBI and IpCOBI retain their original sign convention. The values −IpCOHP($E_F$) and IpCOBI($E_F$) are reported as the bond-averaged −ICOHP and ICOBI (Table S1).

#### 4.1.4 Lattice Thermal Conductivity

The lattice thermal conductivity was evaluated with the WTE[60], which generalizes the linearized BTE by including the off-diagonal elements of the heat-flux operator and yields $\kappa_L = \kappa_{\text{pop}} + \kappa_{\text{coh}}$. The population term, which is the particle-like limit of the WTE, was obtained from the linearized BTE as

$$\kappa_{\text{pop}}^{\alpha\beta} = \frac{1}{k_B T^2 \Omega N} \sum_\lambda n_\lambda^0 (n_\lambda^0 + 1)(\hbar\omega_\lambda)^2 v_\lambda^\alpha F_\lambda^\beta, \tag{4}$$

where $\Omega$ is the unit-cell volume, $N$ is the number of $\mathbf{q}$ points, and $n_\lambda^0$, $\omega_\lambda$, and $\mathbf{v}_\lambda$ are the Bose–Einstein occupation, frequency, and group velocity of mode $\lambda = (\mathbf{q}, s)$ with wavevector $\mathbf{q}$ and branch index $s$. The mean free displacement $\mathbf{F}_\lambda = \tau_\lambda(\mathbf{v}_\lambda + \boldsymbol{\Delta}_\lambda)$ contains the correction $\boldsymbol{\Delta}_\lambda$, which couples mode $\lambda$ to all other modes through the 3ph and 4ph collision operator and was updated self-consistently. Setting $\boldsymbol{\Delta}_\lambda = 0$ recovers the relaxation-time approximation (RTA)[116]. The coherence term, which describes wave-like interband transport between quasi-degenerate branches with overlapping linewidths, was evaluated as

$$\kappa_{\mathrm{coh}}^{\alpha\beta} = \frac{\hbar^2}{k_B T^2 \Omega N} \sum_{\mathbf{q}} \sum_{s \neq s'} \frac{\omega_{\mathbf{q}}^s + \omega_{\mathbf{q}}^{s'}}{2} \, v_{\mathbf{q}}^{s,s',\alpha} \, v_{\mathbf{q}}^{s',s,\beta} \, \frac{\omega_{\mathbf{q}}^s n_{\mathbf{q}}^s (n_{\mathbf{q}}^s + 1) + \omega_{\mathbf{q}}^{s'} n_{\mathbf{q}}^{s'} (n_{\mathbf{q}}^{s'} + 1)}{4(\omega_{\mathbf{q}}^s - \omega_{\mathbf{q}}^{s'})^2 / (\Gamma_{\mathbf{q}}^s + \Gamma_{\mathbf{q}}^{s'}) + (\Gamma_{\mathbf{q}}^s + \Gamma_{\mathbf{q}}^{s'})}, \tag{5}$$

where $\Gamma_{\mathbf{q}}^s = 1/\tau_{\mathbf{q}}^s$ is the total scattering rate of mode $(\mathbf{q}, s)$, and $v_{\mathbf{q}}^{s,s'} = \langle e_{\mathbf{q}}^s | \partial_{\mathbf{q}} D(\mathbf{q}) | e_{\mathbf{q}}^{s'} \rangle / (\omega_{\mathbf{q}}^s + \omega_{\mathbf{q}}^{s'})$ is the velocity-operator element between branches $s$ and $s'$ at the same wavevector[60], with $e_{\mathbf{q}}^s$ the polarization vector and $D(\mathbf{q})$ the dynamical matrix[117].

Population conductivities were obtained from full iterative solutions of the linearized BTE with ShengBTE and FourPhonon[116;118;119], using the TDEP force constants at each temperature. The 3ph-only conductivity, $\kappa_{\mathrm{pop}}^{\mathrm{3ph}}$, was calculated on an $11 \times 11 \times 11$ $\mathbf{q}$ mesh, and the conductivity including 3ph and 4ph scattering, $\kappa_{\mathrm{pop}}^{\mathrm{3+4ph}}$, was calculated on a $7 \times 7 \times 3$ mesh with the complete iterative collision operator. Isotope scattering was included for natural isotopic abundances. Long-range dipole–dipole interactions were included in the dynamical matrix through the high-frequency dielectric tensor and Born effective charges calculated with density functional perturbation theory[65]. In the transport calculations, group velocities were obtained from the $\mathbf{q}$-derivative of the dynamical matrix. The frequency-resolved velocities in Fig. 5(e–h) were evaluated from the same TDEP dispersions with VASPKIT[115]. The mean directional velocity was defined as $\bar{v}_\alpha = \sum_\lambda w_{\mathbf{q}} |v_\lambda^\alpha| / \sum_\lambda w_{\mathbf{q}}$, averaged over all branches with $\mathbf{q}$-point multiplicities $w_{\mathbf{q}}$, and the in-plane mean of $Bi_2O_2S$ was taken as $(\bar{v}_a + \bar{v}_b)/2$. The coherence term was evaluated on a Γ-centered $7 \times 7 \times 3$ mesh with an implementation based on Phonopy and Phono3py, using linewidths that include isotope, 3ph, and 4ph scattering computed on the same mesh. The total conductivity was obtained as $\kappa_L = \kappa_{\mathrm{pop}}^{\mathrm{3+4ph}} + \kappa_{\mathrm{coh}}$.

#### 4.1.5 Homogeneous Nonequilibrium Molecular Dynamics

HNEMD simulations of $Bi_2O_2S$ were performed with GPUMD[105;106] using the same NEP. The 320-atom structural model was replicated 5×5×2, yielding a simulation cell containing 16,000 atoms. The system was equilibrated at the target temperatures (300 and 500 K) and zero pressure in the isothermal–isobaric ensemble for 100 ps with a 1 fs time step. The ensemble was then switched to $NVT$, and a driving force with $F_e = 4\times10^{-4}$ Å$^{-1}$ (4 $\mu$m$^{-1}$), chosen within the linear-response regime, was applied separately along the $a$, $b$, and $c$ axes. The thermal conductivity was obtained from the running average of the induced heat current over the 1.5 ns production trajectory. Reported values and uncertainties are the means and standard errors of five independent simulations for each direction. Because HNEMD samples classical dynamics, it includes anharmonicity to all orders but uses classical mode occupations.

#### 4.1.6 Mode Grüneisen Parameters and Atomic Projections

Mode-resolved volumetric Grüneisen parameters were calculated from the 300 K TDEP force constants on the $7 \times 7 \times 3$ mesh used in the FourPhonon calculations. For mode $\lambda$,

$$\gamma_\lambda = -\frac{\partial \ln \omega_\lambda}{\partial \ln V} = -\frac{V}{\omega_\lambda} \frac{\partial \omega_\lambda}{\partial V}, \tag{6}$$

which was evaluated from the third-order IFCs with the implementation in FourPhonon. The complex phonon eigenvectors were exported with a modified ShengBTE build using the same second-order IFCs, crystal structures, and **q** mesh. The participation of atomic species $s$ in mode $\lambda$ was calculated as

$$P_{s\lambda} = \sum_{\kappa \in s} \sum_{\alpha = x,y,z} |e^{\lambda}_{\kappa\alpha}|^2, \qquad \sum_{s} P_{s\lambda} = 1, \tag{7}$$

where $e^{\lambda}_{\kappa\alpha}$ is the eigenvector of the mass-weighted dynamical matrix normalized as $\sum_{\kappa\alpha} |e^{\lambda}_{\kappa\alpha}|^2 = 1$, and the contributions of the four Bi, two chalcogen, and four O atoms of the 10-atom cell were summed for each species. The heat-capacity-weighted partial Grüneisen parameter of species $s$ within mode group g was evaluated as

$$\bar{\gamma}_{s,g}(T) = \frac{\sum_{\lambda \in g} w_{\mathbf{q}} C_{V,\lambda}(T) \gamma_\lambda P_{s\lambda}}{\sum_{\lambda \in g} w_{\mathbf{q}} C_{V,\lambda}(T)}, \tag{8}$$

where $w_{\mathbf{q}}$ is the symmetry multiplicity of the irreducible **q** point and

$$C_{V,\lambda} = k_B \frac{x_\lambda^2 e^{x_\lambda}}{(e^{x_\lambda} - 1)^2}, \qquad x_\lambda = \frac{\hbar\omega_\lambda}{k_B T}. \tag{9}$$

The partial contributions satisfy $\sum_s \bar{\gamma}_{s,g} = \bar{\gamma}_g$, and the overall average $\bar{\gamma}$ was obtained by extending the sums in eq 8 to all modes. The frequency-ordered branches were divided into the lowest three branches ($A$), branches 4–18 ($O_L$), and the O-dominated branches 19–30 ($O_H$). Because the groups are defined by frequency ordering in the same 10-atom cell, each contains the same fraction of vibrational states in all three compounds (10%, 50%, and 40%, respectively). In the tetragonal compounds, for which the 10-atom cell is the conventional cell, the $A$ group therefore comprises predominantly the lower-frequency half of the primitive-cell acoustic states, and the folded acoustic states at higher frequency fall in $O_L$. All folded branches were included with their associated **q**-point weights.

#### 4.1.7 Thermal Expansion and Heat Capacity

Thermal expansion was calculated by PIMD driven by the NEPs, with classical isothermal–isobaric MD and the QHA used for comparison[61;62]. The GPUMD PIMD integrator applies Langevin thermostats to the centroid and internal ring-polymer modes and a Berendsen barostat for pressure control, whereas the classical simulations used Berendsen temperature and pressure coupling. The PIMD simulations used 48 beads, which exceeds $h\nu_{\max}/k_B T \approx 8$ at 100 K by a factor of six, and a supercell containing 10,240 atoms. Separate simulations were performed at zero external pressure from 100 to 700 K in 50 K steps, each comprising $10^5$ steps of 1 fs, of which the first 70 ps were discarded for equilibration and the remaining 30 ps were used for structural averaging. The classical simulations used the same supercell and time step with $10^6$ steps per temperature, of which 700 ps were used for equilibration and 300 ps for averaging. Lattice parameters and cell volumes were sampled every 0.1 ps.

To separate the roles of atomic mass and bonding, additional PIMD simulations of $Bi_2O_2Se$ were performed with the Se mass (78.97 amu) replaced by the S mass (32.06 amu), while the $Bi_2O_2Se$ NEP, supercell, and simulation protocol, including five independent trajectories, were retained. Because the potential energy surface is unchanged, any difference relative to $Bi_2O_2Se$ arises solely from the mass dependence of nuclear quantum effects.

The linear and volumetric CTEs were calculated as

$$\alpha_i(T) = \frac{1}{L_i(T)} \left( \frac{\partial L_i}{\partial T} \right)_P, \qquad \alpha_V(T) = \frac{1}{V(T)} \left( \frac{\partial V}{\partial T} \right)_P, \tag{10}$$

where $L_i$ is the lattice parameter along axis $i$ and $V$ is the cell volume. The derivatives were evaluated by central finite differences with a 50 K spacing from 150 to 650 K and by second-order forward and backward differences at 100 and 700 K, respectively, yielding tangent coefficients at each temperature. Five independent simulations were performed for each condition, and the reported CTEs and uncertainties are the means and 95% confidence intervals of the five estimates.

The constant-pressure heat capacity of $Bi_2O_2Se$ used in the FDTR analysis was obtained from the PIMD simulations by differentiating a cubic polynomial fit to the temperature-dependent mean enthalpy and was converted to a volumetric heat capacity with the PIMD density. Because PIMD includes nuclear quantum effects, this heat capacity captures the suppressed contribution of the high-frequency O modes and agrees with reported measurements (Fig. S5).

For the QHA, the temperature-dependent equilibrium volume was obtained by minimizing the Helmholtz free energy $F(V,T) = E_0(V) + F_{\text{vib}}(V,T)$ over 11 volumes spanning $-1.5\%$ to $+1.5\%$ of the reference volume. The harmonic force constants were constructed from DFT forces, and the vibrational free energies were evaluated with Phonopy[65;120;121]. Because the QHA was restricted to isotropic volume changes, it yields only $\alpha_V$, which was compared with the PIMD and classical MD values to assess the role of explicit anharmonicity.

#### 4.1.8 Device-Scale Phonon Transport and Thermoelastic Simulations

Steady-state phonon transport in a three-dimensional FinFET channel was simulated with the non-gray phonon BTE solver GiftBTE[20;21], which accounts for ballistic transport and phonon–boundary scattering. GiftBTE solves the BTE within the RTA using the discrete ordinates method. The mode-resolved frequencies, group velocities, heat capacities, and relaxation times of each compound were taken from the ShengBTE and FourPhonon calculations described above, with scattering rates including isotope, 3ph, and 4ph processes. The angular space was discretized into 16 directions and the phonon properties into 15 mean-free-path bands. The BTE describes particle-like transport and therefore omits the coherence contribution. Both the normal- and elevated-load simulations used property sets evaluated at 300 K, and additional elevated-load simulations with property sets evaluated at 500 and 700 K were used to assess the sensitivity to the property-evaluation temperature (Table S7).

The device geometry and coordinate convention were adapted from previous FinFET electrothermal models and the GiftBTE FinFET benchmark[21;122]. The $x$, $y$, and $z$ directions correspond to the channel, fin-width, and fin-height directions, respectively. The fin had a width $W = 12$ nm, a height $H = 60$ nm, and a length $L = 50$ nm. Following the vertical-layer orientation of fabricated $Bi_2O_2Se$ FinFETs[11;14], the crystallographic $a$, $c$, and $b$ axes were aligned with $x$, $y$, and $z$, respectively, so that the $[Bi_2O_2]^{2+}$ layers lie parallel to the sidewalls. For tetragonal $Bi_2O_2Se$ and $Bi_2O_2Te$, the $a$ and $b$ axes are equivalent. In the thermoelastic model, a 5.5 nm thick $HfO_2$ gate dielectric covered the top and both sidewalls within the gate region, and the fin rested on a 30 nm thick $LaAlO_3$ substrate. The source and drain were represented by boundary conditions at the channel end faces.

A semi-elliptical cylindrical heat-generation region spanning the fin width was placed beneath the top surface near the drain. Its cross-section lay in the $x$–$z$ plane, with a channel-direction semi-axis of 10 nm and a depth semi-axis of 15 nm, and its center was located on the top surface at $x/L = 0.26$, with $x$ measured from the channel end adjacent to the hotspot. Uniform volumetric heat-generation rates of $Q_1 = 1.4 \times 10^{18}$ and $Q_2 = 5.8 \times 10^{18}\ \text{W m}^{-3}$, corresponding to total powers of approximately 3.94 and 16.34 $\mu$W per fin based on the discretized heat-source volume, were applied for the normal- and elevated-load cases, respectively. The generated heat was distributed among phonon modes in proportion to their modal heat capacities[21;123].

For the thermal calculations, the fin and supporting base were treated using the same phonon-property set, with the bottom of the base maintained at 300 K. Thermalizing boundaries at 300 K were also applied to the channel faces adjoining the source and drain contacts[21;122]. The fin–dielectric interfaces were represented by adiabatic, diffusely reflecting phonon boundaries, following the GiftBTE FinFET benchmark[21]. The lateral faces of the base were treated as symmetry planes between neigh-boring fins[21;123–125]. For the Fourier benchmark, the fin and supporting base were assigned the bulk conductivity tensor computed within the RTA from the same modal properties, and the thermal geometry, mesh, material properties, heat source, and boundary conditions were kept identical, so that the BTE and Fourier models share the same bulk limit. The same unstructured mesh was used for the BTE and Fourier calculations. Thermal and thermoelastic meshes contained 16,930–125,289 and 22,370–157,457 nodes, respectively. Relative to the finest-mesh results, the maximum deviations were 0.41% in $T_{\mathrm{hot}}$ and 4.41% in $\varepsilon_{\mathrm{mis}}^{\mathrm{max}}$ (Table S6).

A sequential, one-way thermal–mechanical approach was adopted. The calculated temperature fields were transferred to the COMSOL Multiphysics thermoelastic mesh, which additionally included the gate dielectric as a bonded elastic domain to account for elastic constraint and thermal-expansion mismatch. The dielectric temperature was prescribed from the adjacent fin surface and assumed uniform through its thickness. This approximation neglects heat transport through the gate stack and deformation-induced changes in thermal transport.

Stationary, small-strain linear-elastic calculations were performed using these temperature fields as thermal loads. The channel was described by the anisotropic elastic stiffness tensor of each compound, taken from the first-principles calculations of Hu et al.[66], whereas the dielectric and substrate were treated as isotropic solids with the Young's moduli, Poisson's ratios, and CTEs listed in Table S8. Displacement continuity was enforced at all interfaces. The bottom face of the substrate was constrained in the normal direction, its lateral faces were assigned symmetry (roller) conditions, and all remaining surfaces were traction-free. The temperature-dependent PIMD coefficients $\alpha_a(T)$, $\alpha_b(T)$, and $\alpha_c(T)$ were supplied as tangent CTEs, so that the thermal strain along each crystallographic axis relative to the stress-free reference temperature of 300 K was

$$\varepsilon_i^{\mathrm{th}}(\mathbf{r}) = \int_{300\,\mathrm{K}}^{T(\mathbf{r})} \alpha_i(T')\,dT', \qquad i = a, b, c. \tag{11}$$

The thermal-mismatch strain was defined as the elastic part of the total strain, $\boldsymbol{\varepsilon}_{\mathrm{mis}} = \boldsymbol{\varepsilon} - \boldsymbol{\varepsilon}^{\mathrm{th}}$, which vanishes for a uniformly heated, unconstrained solid. The reported $\varepsilon_{\mathrm{mis}}^{\mathrm{max}}$ is the spatial maximum within the channel of the largest-magnitude principal value of $\boldsymbol{\varepsilon}_{\mathrm{mis}}$.

## 4.2 Experimental Methods

### 4.2.1 Materials

Millimeter-scale $Bi_2O_2Se$ single crystals were obtained from SixCarbon Technology and used for Raman spectroscopy, temperature-dependent XRD, and FDTR measurements.

### 4.2.2 Raman Spectroscopy

Room-temperature Raman spectra were collected from the (001) surface of a $Bi_2O_2Se$ crystal before the thermal measurements with a Renishaw inVia micro-Raman system using 514 nm excitation at 2.5 mW. Spectra were acquired from 100 to 500 $\mathrm{cm}^{-1}$ through a 50× objective with three accumulations of 60 s each.

### 4.2.3 Temperature-Dependent X-ray Diffraction

Temperature-dependent XRD was performed on a $Bi_2O_2Se$ crystal with lateral dimensions of about 3 mm × 3 mm and a thickness above 50 $\mu$m, supported on a silicon wafer holder to suppress parasitic scattering from the heating stage. Measurements were carried out on a Malvern Panalytical Empyrean diffractometer equipped with an Anton Paar HTK 1200N high-temperature chamber, using Cu $K_\alpha$ radiation ($\lambda_{K\alpha1} = 1.5406$ Å, 45 kV, 40 mA). Patterns were collected in ambient air from 298 to 473 K in 25 K steps over $2\theta = 10°–65°$ with a step size of 0.0066°. The room-temperature pattern was consistent with the tetragonal $I4/mmm$ structure (No. 139), and a pattern recorded after cooling to room temperature confirmed that heating caused no structural degradation. At each temperature, the $c$ lattice parameter was calculated from the positions of the (00$l$) reflections and their Miller indices using Bragg's law. The cross-plane CTE was obtained from a linear fit of $c(T)$ between 298 and 398 K as $\alpha_c = (1/c_{298\,\mathrm{K}})(dc/dT)$, which corresponds to the calculated $\alpha_c$ averaged over the same interval, and its uncertainty was derived from the uncertainty of the fitted slope.

#### 4.2.4 Frequency-Domain Thermoreflectance

The anisotropic thermal conductivity of $Bi_2O_2Se$ was measured by FDTR following our previous work[126;127]. A 445 nm pump laser, intensity-modulated from 0.5 to 10 MHz, and a continuous-wave 532 nm probe laser were focused onto the sample through a 20× objective. The $1/e^2$ spot radius, determined by beam-offset scans, was about 1.94 $\mu$m[128]. Pump and probe powers of 1.5 and 2 mW kept the steady-state temperature rise below about 15 K. For measurements from 300 to 500 K, the sample was mounted on a temperature-controlled heating stage, and data were acquired after the temperature had stabilized at each set point.

A 61.5 nm Pt transducer was sputtered onto the freshly cleaved (001) surface of a bulk crystal with a thickness above 200 $\mu$m. Pt was chosen instead of the more common Au because its lower thermal conductivity ($\kappa_{\mathrm{Pt}} \approx 36.9\,\mathrm{W\,m^{-1}\,K^{-1}}$ at 300 K) limits lateral spreading within the transducer and increases the sensitivity to $\kappa_{\mathrm{in}}$. The Pt thickness was measured with an Alpha-Step D300 profilometer (KLA-Tencor) on a Si reference wafer coated in the same deposition run. $\kappa_{\mathrm{Pt}}$ was calibrated at each measurement temperature by FDTR on a reference sample in which the same Pt film was deposited on a 2 $\mu$m $SiO_2$ layer, and the calibrated values were fixed in the analysis of $Bi_2O_2Se$. The pump–probe phase lag was recorded with a balanced detector and demodulated with a lock-in amplifier (HF2LI, Zurich Instruments). At each temperature, measurements were performed at three locations on the sample surface, each scanned three times.

The phase data were analyzed with a transfer-matrix solution of anisotropic heat diffusion in a multilayer[127]. Because the sensitivities to $\kappa_{\mathrm{in}}$ and $\kappa_{\mathrm{cross}}$ have distinct frequency dependences over 0.5–10 MHz (Fig. S4), $\kappa_{\mathrm{in}}$, $\kappa_{\mathrm{cross}}$, and the Pt/$Bi_2O_2Se$ thermal boundary conductance were fitted simultaneously to the complete phase spectrum. The volumetric heat capacity of $Bi_2O_2Se$ at each temperature was taken from PIMD (Section 4.1.7). All fixed parameters are listed in Table S2. The uncertainties of the fitted conductivities were evaluated by error propagation, assuming a 3% uncertainty in each fixed parameter.

Because FDTR measures the total thermal conductivity, the electronic contribution was estimated from the Wiedemann–Franz law, $\kappa_e = L\sigma T$. We used the in-plane electrical conductivity reported for nominally undoped bulk $Bi_2O_2Se$ single crystals, $\sigma = 8.1 \times 10^3\,\mathrm{S\,m^{-1}}$ at 300 K[48], and approximated its temperature dependence above room temperature as $\sigma \propto T^{-1}$ owing to phonon-limited carrier scattering. A Lorenz number of $L = 1.5 \times 10^{-8}\,\mathrm{W\,\Omega\,K^{-2}}$, representative of a nondegenerate semiconductor dominated by acoustic-phonon scattering, was used[49]. The resulting estimate, $\kappa_e \approx 0.04\,\mathrm{W\,m^{-1}\,K^{-1}}$ over 300–500 K, corresponds to less than 4% of the measured in-plane conductivity, and the cross-plane contribution is smaller still because of the anisotropic electrical conductivity. The electronic contribution was therefore neglected, and the measured values were compared directly with the calculated $\kappa_L$.

## Acknowledgements

This work was supported by the National Key R&D Program of China (Grant No. 2024YFB4405700), the Research Grants Council of Hong Kong (Project No. 16214626), and the National Natural Science Foundation of China (Grant No. 52676057). It was also funded by the Guangdong Basic and Applied Basic Research Foundation (Grant Nos. 2025A1515012898 and 2026A1515011290), the Opening Project of the State Key Laboratory of Advanced Glass Materials (Grant No. 2025KF27), and the Frontier Technology Research for Joint Institutes with Industry Scheme sponsored by the Center on Smart Sensors and Environmental Technologies at HKUST. Furthermore, the authors gratefully acknowledge the HKUST Fok Ying Tung Research Institute and the National Supercomputing Center in Guangzhou (Nansha Sub-center) for providing the high-performance computational resources used in this study.

## Conflicts of Interest

The authors declare no competing financial interest.

## Data Availability

The data supporting this study are available from the corresponding author on reasonable request.

# Supporting Information

## Lone-Pair Stereoactivity Overrides Mass Scaling of Thermal Expansion and Heat Conduction in Non-van der Waals Layered Bismuth Oxychalcogenides

Linzhen Li[1,2,†], Guangwu Zhang[2,†], Ziyan Qian[2,†], Wei Jing[2], Yang Lu[2], Jiaqi Gu[2], Kaifeng Huang[2], Yanguang Zhou[2], and Qiye Zheng[2,*]

[1]School of Energy Engineering, Zhejiang University, Hangzhou 310027, China

[2]Department of Mechanical and Aerospace Engineering, The Hong Kong University of Science and Technology, Hong Kong SAR, China

[†]These authors contributed equally.

[*]Corresponding author: qiyezheng@ust.hk

# I Machine-learning potential validation

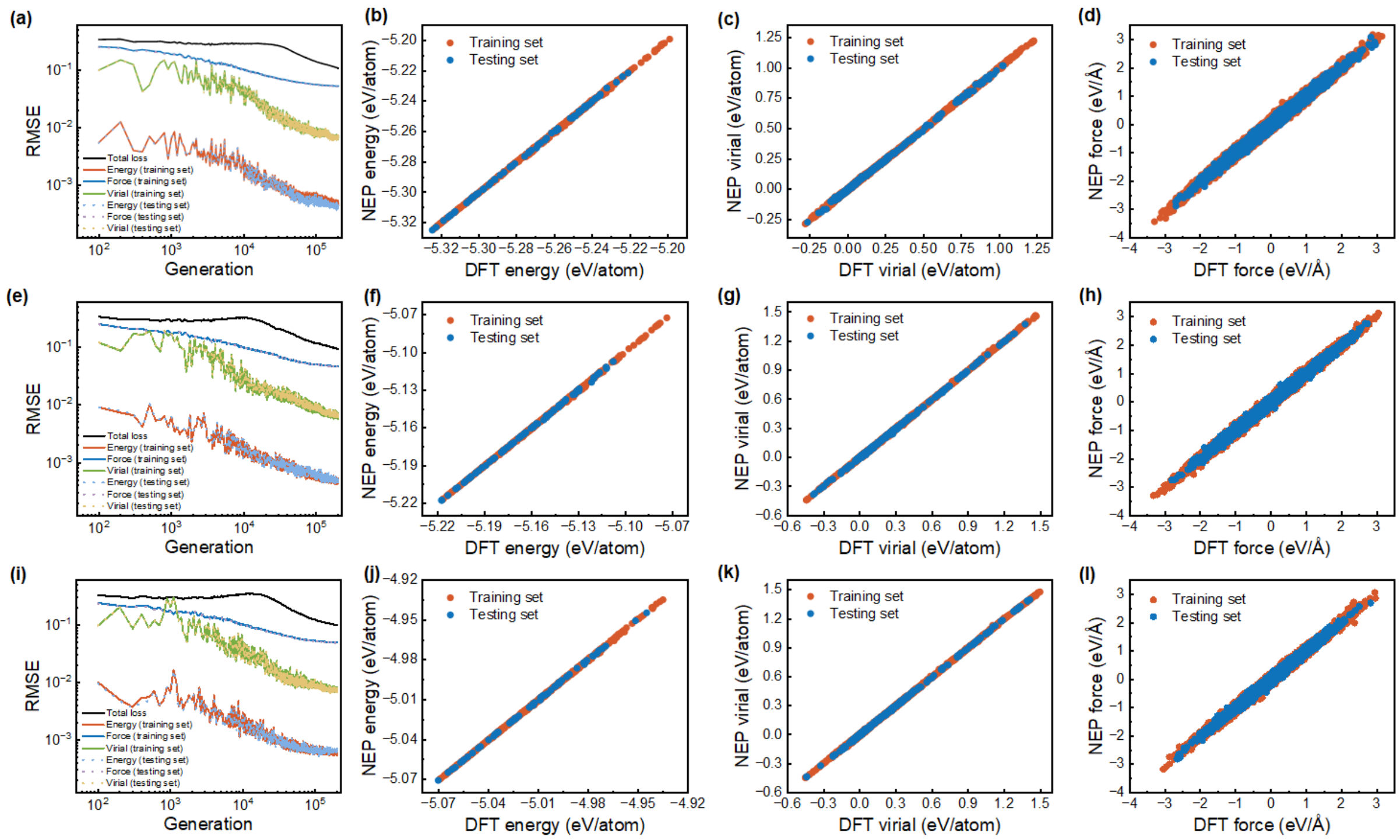

**Figure S1:** Training and validation performance of the NEP model for $Bi_2O_2X$ (X = S, Se, and Te). Panels (a–d), (e–h), and (i–l) present the results for $Bi_2O_2S$, $Bi_2O_2Se$, and $Bi_2O_2Te$, respectively. Panels (a, e, i) show the evolution of the total loss and the root mean square errors (RMSEs) of energy, force, and virial for the training and testing datasets with training generation. Panels (b, f, j), (c, g, k), and (d, h, l) compare the NEP-predicted energies, virials, and forces with the corresponding DFT reference values, respectively. The close agreement between the NEP predictions and DFT data for both datasets demonstrates the accuracy of the trained models.

# II Bond-resolved chemical-bonding metrics

**Table S1:** Bond-averaged −ICOHP and ICOBI values for the nearest-neighbor Bi–O and Bi–X interactions in the fully relaxed $Bi_2O_2X$ structures. The values were averaged over 16 Bi–O and 16 Bi–X interactions in the 10-atom conventional cell. −ICOHP is reported in eV per bond, whereas ICOBI is dimensionless.

| Compound | −ICOHP(Bi–O) (eV bond$^{-1}$) | ICOBI(Bi–O) | −ICOHP(Bi–X) (eV bond$^{-1}$) | ICOBI(Bi–X) |
|---|---|---|---|---|
| $Bi_2O_2S$ | 2.322 | 0.391 | 0.808 | 0.189 |
| $Bi_2O_2Se$ | 2.296 | 0.391 | 0.747 | 0.191 |
| $Bi_2O_2Te$ | 2.186 | 0.386 | 0.769 | 0.203 |

# III QHA benchmark for volumetric thermal expansion

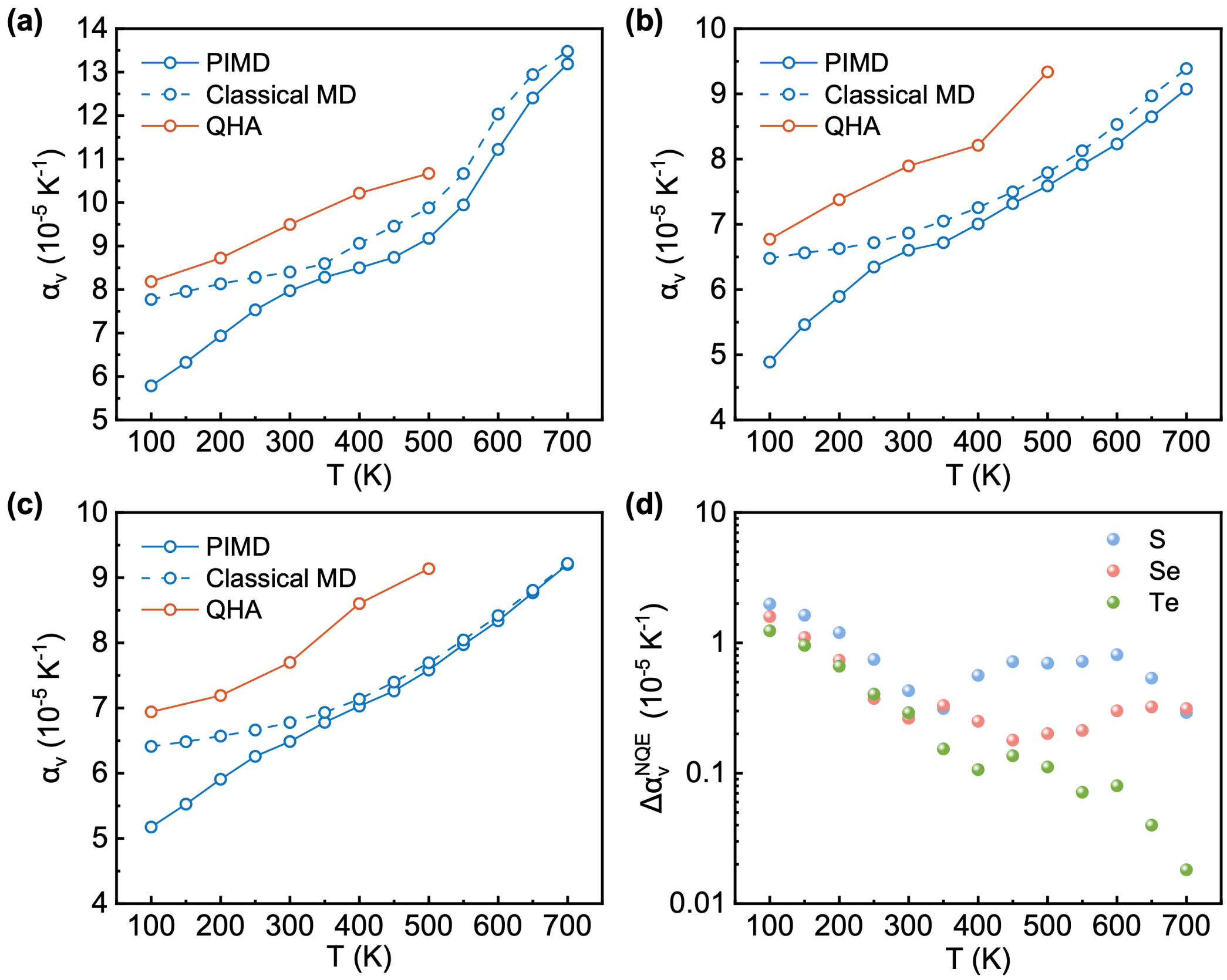

**Figure S2:** Temperature-dependent volumetric coefficients of thermal expansion, $\alpha_V$, of (a) $Bi_2O_2S$, (b) $Bi_2O_2Se$, and (c) $Bi_2O_2Te$ obtained from PIMD, classical NPT-MD, and the quasi-harmonic approximation (QHA). (d) Magnitude of the NQE-induced correction, defined as $\Delta\alpha_V^{\mathrm{NQE}} = \alpha_V^{\mathrm{classical\ MD}} - \alpha_V^{\mathrm{PIMD}}$, for the three compounds.

## IV Raman characterization of $Bi_2O_2Se$

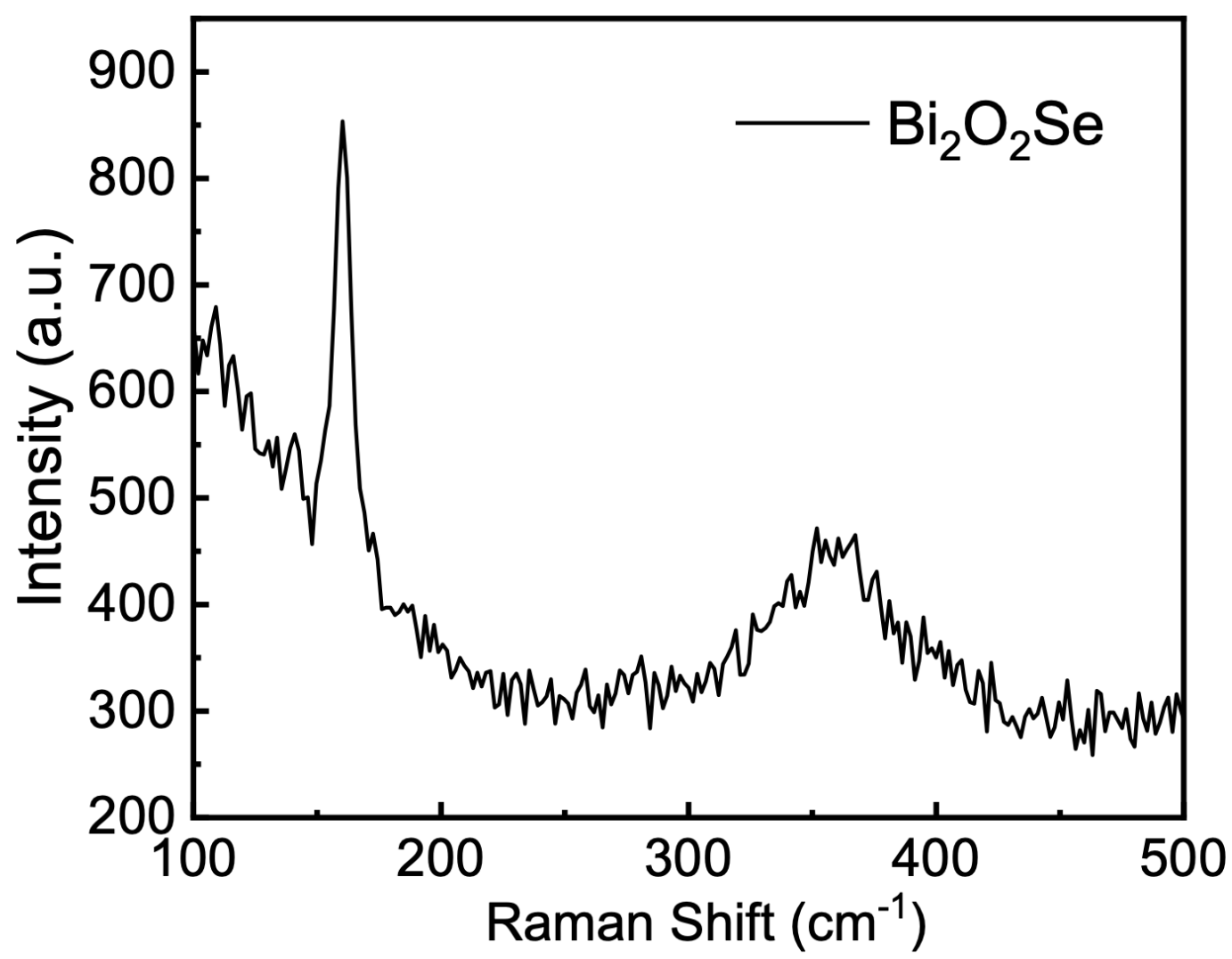

**Figure S3:** Room-temperature Raman spectrum of the $Bi_2O_2Se$ single crystal measured from the (001) surface. The prominent peak near 160 $cm^{-1}$ and the weaker broad feature at 350–365 $cm^{-1}$ are assigned to the $A_{1g}$ and $B_{1g}$ modes, respectively, consistent with tetragonal $Bi_2O_2Se$.

# V FDTR parameters and sensitivity analysis

**Table S2:** Material properties in the FDTR fitting model.

| Layer | $C_p$ (J cm$^{-3}$ K$^{-1}$) | $\kappa_{\perp}$ (W m$^{-1}$ K$^{-1}$) | $\kappa_{\parallel}$ (W m$^{-1}$ K$^{-1}$) | Thickness (nm) |
|---|---|---|---|---|
| Pt | 2.85 [1] | 36.9 | 36.9 | 61.5 |
| G1 | – | fitted | – | – |
| $Bi_2O_2Se$ | 1.96 | fitted | fitted | infinite |

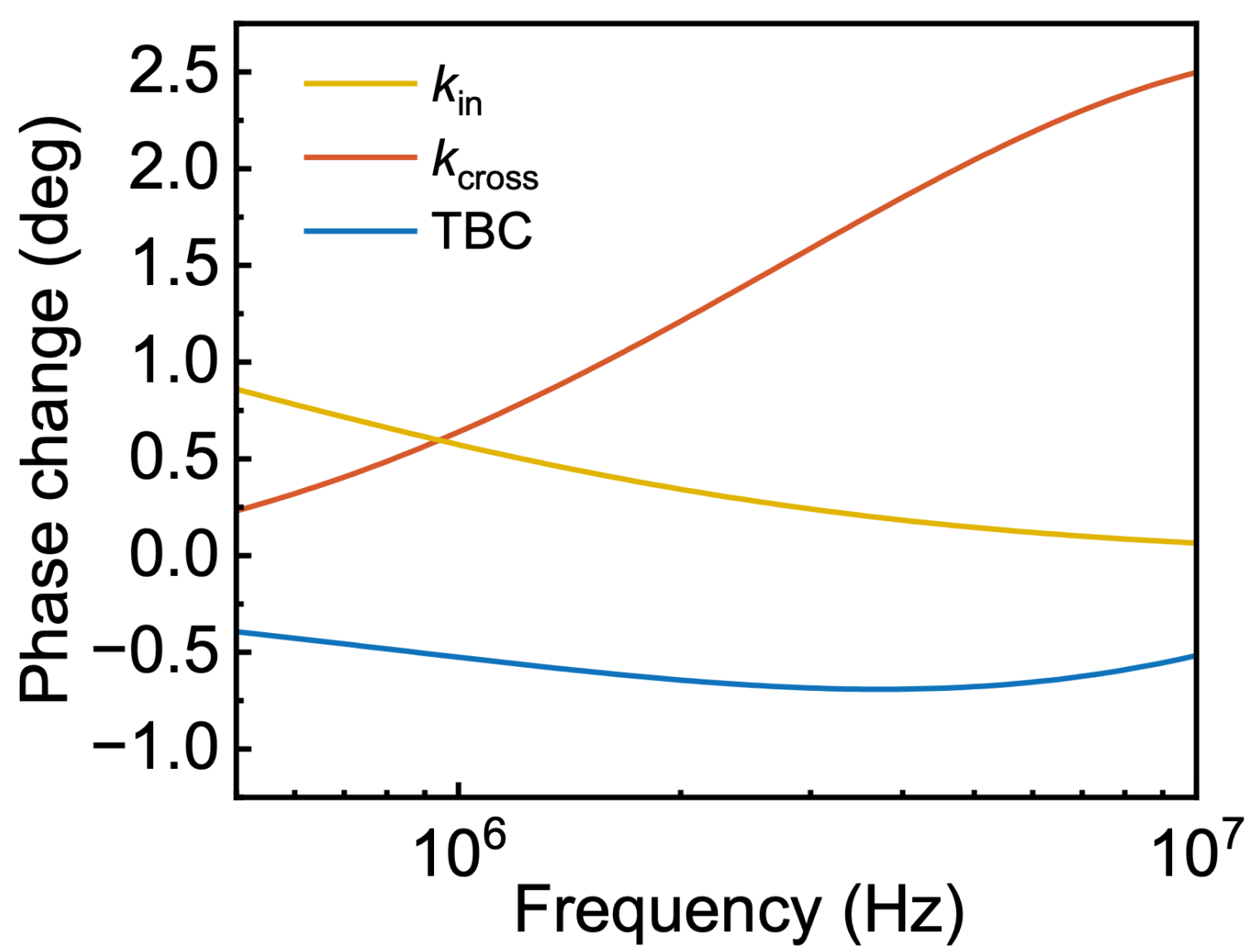

**Figure S4:** Frequency-dependent sensitivity of the FDTR phase response of $Bi_2O_2Se$ to the in-plane thermal conductivity $\kappa_{in}$, cross-plane thermal conductivity $\kappa_{cross}$, and thermal boundary conductance (TBC). The distinct frequency dependences of $\kappa_{in}$ and $\kappa_{cross}$ enable their simultaneous determination from the same phase dataset.

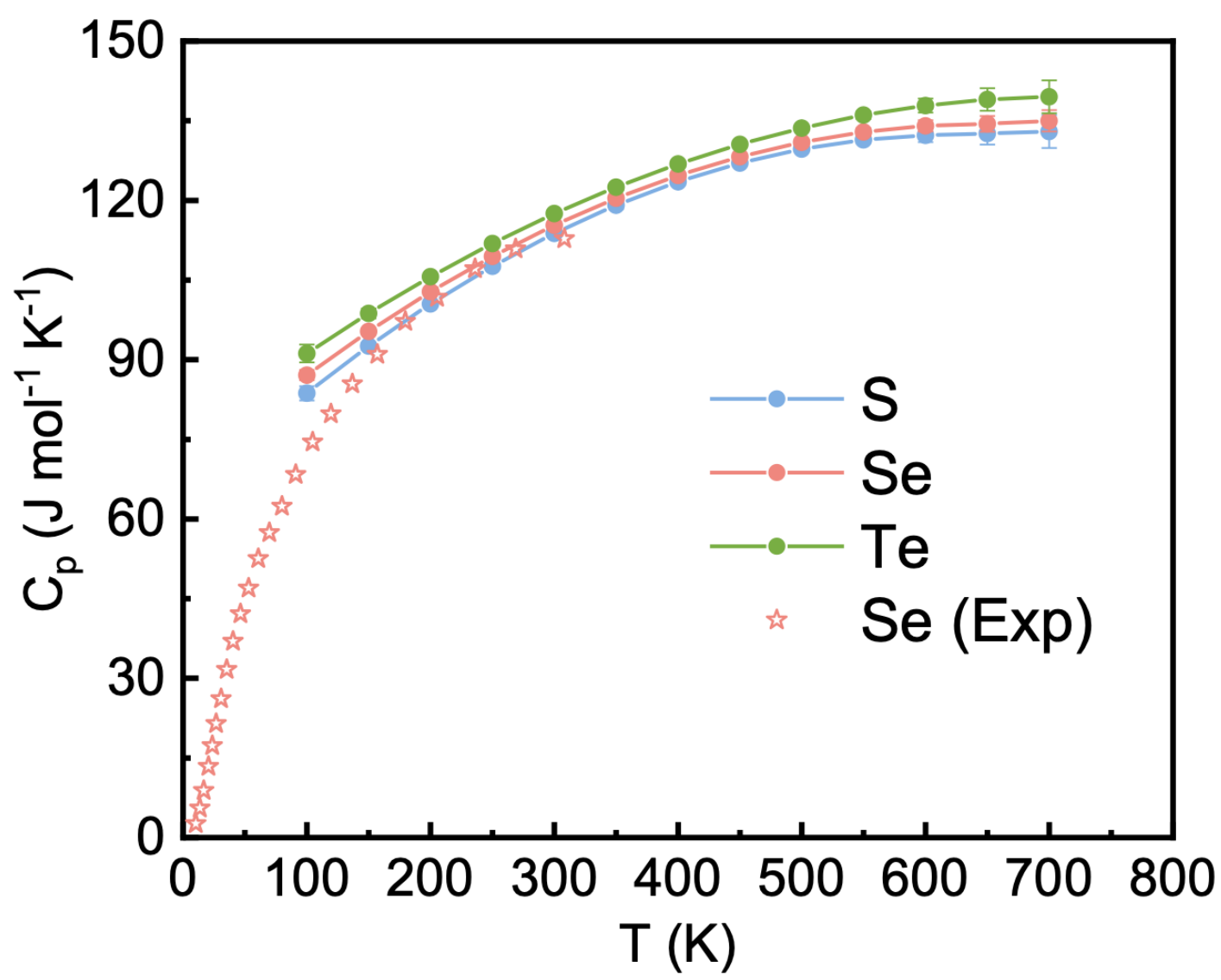

**Figure S5:** Temperature-dependent molar heat capacities at constant pressure, $C_p$, of $Bi_2O_2S$, $Bi_2O_2Se$, and $Bi_2O_2Te$ obtained from the temperature derivatives of cubic-polynomial fits to the mean enthalpies from NEP-driven GPUMD simulations. Open stars denote the experimental values for $Bi_2O_2Se$ reported by Zich et al. [2].

## VI Detailed comparison of calculated CTEs

**TABLE S3.** Thermal-expansion data at 300 K. (a) In-plane and cross-plane coefficients of thermal expansion of $Bi_2O_2S$, $Bi_2O_2Se$, and $Bi_2O_2Te$ obtained from classical MD and PIMD, together with the experimentally measured cross-plane CTE of $Bi_2O_2Se$. (b) Volumetric CTEs of $Bi_2O_2X$ and the literature S–Se–Te chalcogenide series used in Fig. 3(b), together with the corresponding sulfide-normalized ratios and their assignment to the shaded envelope representing the prevailing trend.

### (a) Direction-resolved CTEs of $Bi_2O_2X$.

| Material | $\alpha_{in}$ (MD) | $\alpha_{cross}$ (MD) | $\alpha_{in}$ (PIMD) | $\alpha_{cross}$ (PIMD) | $\alpha_{cross}$ (Exp.) |
|---|---|---|---|---|---|
| $Bi_2O_2S$ | $1.93^a$, $3.11^b$ | 3.20 | $1.89^a$, $3.08^b$ | 3.17 | – |
| $Bi_2O_2Se$ | 2.15 | 2.75 | 2.04 | 2.64 | 2.82 |
| $Bi_2O_2Te$ | 2.03 | 2.74 | 1.97 | 2.69 | – |

CTEs are given in $10^{-5}$ $K^{-1}$ at 300 K. Superscripts $a$ and $b$ denote the crystallographic $a$ and $b$ directions of orthorhombic $Bi_2O_2S$, respectively. Replacing the Se atomic mass with the S atomic mass while retaining the $Bi_2O_2Se$ potential changes its PIMD volumetric CTE by only 0.6% at 300 K.

## (b) Source data for the sulfide-normalized volumetric-CTE comparison in Fig. 3(b).

| Material series | $\alpha_V$(S) | $\alpha_V$(Se) | $\alpha_V$(Te) | $\alpha_V$(Se)/$\alpha_V$(S) | $\alpha_V$(Te)/$\alpha_V$(S) | Included in shaded envelope | Source |
|---|---|---|---|---|---|---|---|
| $Bi_2O_2X$ | 8.14 | 6.72 | 6.61 | 0.826 | 0.812 | No—this work | This work |
| CdX$^{\dagger}$ | 1.512 | 1.733 | 1.993 | 1.146 | 1.318 | Yes | [3] |
| CaX* | 5.07 | 6.03 | 6.12 | 1.189 | 1.207 | Yes | [4] |
| $Sr_{0.5}Ca_{0.5}X$* | 5.81 | 6.04 | 6.36 | 1.040 | 1.094 | Yes | [4] |
| $CuTlX_2$ | 4.87 | 5.04 | 5.10 | 1.035 | 1.047 | Yes | [5] |
| NiX* | 3.84 | 5.00 | 4.74 | 1.302 | 1.234 | Yes | [6] |
| $Cu_2MgSnX_4$* | 6.09 | 6.70 | 6.99 | 1.100 | 1.148 | Yes | [7] |
| YX, B1 phase | 3.70 | 4.10 | 4.30 | 1.108 | 1.162 | Yes | [8] |
| $CuInX_2$* | 4.67 | 4.81 | 5.18 | 1.030 | 1.109 | Yes | [9] |
| ZnX$^{\dagger}$ | 2.625 | 1.965 | 2.010 | 0.749 | 0.766 | No | [10] |
| $AgGaX_2$* | 4.81 | 4.72 | 4.79 | 0.981 | 0.996 | No | [11] |

Volumetric CTEs are given in $10^{-5}$ $K^{-1}$ at 300 K and ambient pressure. The normalized ratios were calculated from the values before rounding. The shaded envelope in Fig. 3(b) comprises the series for which both $\alpha_V(\mathrm{Se})/\alpha_V(\mathrm{S})$ and $\alpha_V(\mathrm{Te})/\alpha_V(\mathrm{S})$ exceed unity. $Bi_2O_2X$, ZnX, and $AgGaX_2$ are therefore not included in the envelope and are shown separately. *Values digitized from the corresponding published curves at 300 K and 0 GPa. $^{\dagger}$For cubic materials reported in terms of linear CTE, $\alpha_V = 3\alpha_L$.

## VII Higher-order phonon scattering and temperature scaling

**TABLE S4.** Temperature-dependent Peierls population conductivities calculated with three-phonon scattering only ($\kappa_{p,i}^{3ph}$) and with combined three- and four-phonon scattering ($\kappa_{p,i}^{3+4ph}$), together with the Wigner coherence contribution ($\kappa_{c,i}$), for (a) $Bi_2O_2S$, (b) $Bi_2O_2Se$, and (c) $Bi_2O_2Te$. All values are given in W m$^{-1}$ K$^{-1}$.

**(a) $Bi_2O_2S$.**

| $T$ (K) | $\kappa_{p,a}^{3ph}$ | $\kappa_{p,a}^{3+4ph}$ | $\kappa_{c,a}$ | $\kappa_{p,b}^{3ph}$ | $\kappa_{p,b}^{3+4ph}$ | $\kappa_{c,b}$ | $\kappa_{p,cross}^{3ph}$ | $\kappa_{p,cross}^{3+4ph}$ | $\kappa_{c,cross}$ |
|---|---|---|---|---|---|---|---|---|---|
| 100 | 3.1024 | 2.4957 | 0.1750 | 4.4254 | 2.7847 | 0.1183 | 3.6537 | 1.1721 | 0.1359 |
| 200 | 1.7897 | 1.2705 | 0.2057 | 2.2681 | 1.3096 | 0.1667 | 1.6522 | 0.5329 | 0.1793 |
| 300 | 1.1821 | 0.8339 | 0.2097 | 2.1431 | 0.9043 | 0.1735 | 0.9467 | 0.3328 | 0.1741 |
| 400 | 1.0010 | 0.5788 | 0.2158 | 1.1892 | 0.5689 | 0.1944 | 0.6136 | 0.1977 | 0.1911 |
| 500 | 0.4550 | 0.3282 | 0.1931 | 0.5023 | 0.3260 | 0.1846 | 0.3413 | 0.1198 | 0.1670 |
| 600 | 0.2998 | 0.2168 | 0.1684 | 0.3407 | 0.2252 | 0.1641 | 0.1403 | 0.0617 | 0.1633 |
| 700 | 0.2515 | 0.1565 | 0.1682 | 0.3157 | 0.1793 | 0.1547 | 0.0942 | 0.0433 | 0.1578 |

**(b) $Bi_2O_2Se$.**

| $T$ (K) | $\kappa_{p,in}^{3ph}$ | $\kappa_{p,in}^{3+4ph}$ | $\kappa_{c,in}$ | $\kappa_{p,cross}^{3ph}$ | $\kappa_{p,cross}^{3+4ph}$ | $\kappa_{c,cross}$ |
|---|---|---|---|---|---|---|
| 100 | 6.3641 | 5.5118 | 0.1399 | 5.9833 | 2.9518 | 0.0378 |
| 200 | 3.2882 | 2.7134 | 0.1401 | 2.6062 | 1.3733 | 0.0568 |
| 300 | 2.2091 | 1.6887 | 0.1439 | 1.6430 | 0.8741 | 0.0736 |
| 400 | 1.7702 | 1.3152 | 0.1385 | 1.2614 | 0.7159 | 0.0795 |
| 500 | 1.4371 | 1.0029 | 0.1442 | 0.9108 | 0.4979 | 0.0930 |
| 600 | 1.2965 | 0.8841 | 0.1412 | 0.7666 | 0.4630 | 0.0900 |
| 700 | 1.1322 | 0.7764 | 0.1384 | 0.6258 | 0.4188 | 0.0957 |

**(c) $Bi_2O_2Te$.**

| $T$ (K) | $\kappa_{p,in}^{3ph}$ | $\kappa_{p,in}^{3+4ph}$ | $\kappa_{c,in}$ | $\kappa_{p,cross}^{3ph}$ | $\kappa_{p,cross}^{3+4ph}$ | $\kappa_{c,cross}$ |
|---|---|---|---|---|---|---|
| 100 | 5.9847 | 3.8578 | 0.0974 | 1.6608 | 1.0593 | 0.0260 |
| 200 | 3.1327 | 2.3818 | 0.0981 | 0.9063 | 0.6190 | 0.0340 |
| 300 | 2.1726 | 1.6426 | 0.1117 | 0.6355 | 0.4614 | 0.0475 |
| 400 | 1.6752 | 1.2597 | 0.1132 | 0.5177 | 0.3484 | 0.0507 |
| 500 | 1.4799 | 1.0909 | 0.1030 | 0.4154 | 0.2932 | 0.0500 |
| 600 | 1.3125 | 0.9689 | 0.1022 | 0.3311 | 0.2322 | 0.0530 |
| 700 | 1.2051 | 0.8013 | 0.1038 | 0.2974 | 0.1971 | 0.0525 |

**TABLE S5.** Temperature dependence of the four-phonon-induced reduction ($R_{4\text{ph},i}$) and Wigner coherence fraction ($f_{c,i}$) of $Bi_2O_2S$, $Bi_2O_2Se$, and $Bi_2O_2Te$. The four-phonon-induced reduction and coherence fraction, both expressed as percentages, are defined as $R_{4\text{ph},i} = (\kappa_{p,i}^{3\text{ph}} - \kappa_{p,i}^{3+4\text{ph}})/\kappa_{p,i}^{3\text{ph}} \times 100\%$ and $f_{c,i} = \kappa_{c,i}/(\kappa_{p,i}^{3+4\text{ph}} + \kappa_{c,i}) \times 100\%$. Values were calculated from the unrounded conductivities reported in Table S4.

**(a) $Bi_2O_2S$.**

| $T$ (K) | $R_{4\text{ph},a}$ | $R_{4\text{ph},b}$ | $R_{4\text{ph,cross}}$ | $f_{c,a}$ | $f_{c,b}$ | $f_{c,\text{cross}}$ |
|---|---|---|---|---|---|---|
| 100 | 19.6 | 37.1 | 67.9 | 6.6 | 4.1 | 10.4 |
| 200 | 29.0 | 42.3 | 67.7 | 13.9 | 11.3 | 25.2 |
| 300 | 29.5 | 57.8 | 64.8 | 20.1 | 16.1 | 34.3 |
| 400 | 42.1 | 52.2 | 67.8 | 27.2 | 25.5 | 49.2 |
| 500 | 27.9 | 35.1 | 64.9 | 37.0 | 36.1 | 58.2 |
| 600 | 27.7 | 33.9 | 56.0 | 43.7 | 42.2 | 72.6 |
| 700 | 37.8 | 43.2 | 54.0 | 51.8 | 46.3 | 78.5 |

**(b) $Bi_2O_2Se$.**

| $T$ (K) | $R_{4\text{ph,in}}$ | $R_{4\text{ph,cross}}$ | $f_{c,\text{in}}$ | $f_{c,\text{cross}}$ |
|---|---|---|---|---|
| 100 | 13.4 | 49.4 | 2.5 | 1.3 |
| 200 | 17.5 | 47.3 | 4.9 | 4.0 |
| 300 | 23.6 | 46.8 | 7.9 | 7.8 |
| 400 | 25.7 | 43.2 | 9.5 | 10.0 |
| 500 | 30.2 | 49.1 | 12.6 | 15.7 |
| 600 | 31.8 | 39.6 | 13.8 | 16.3 |
| 700 | 31.4 | 33.1 | 15.1 | 18.6 |

**(c) $Bi_2O_2Te$.**

| $T$ (K) | $R_{4\text{ph,in}}$ | $R_{4\text{ph,cross}}$ | $f_{c,\text{in}}$ | $f_{c,\text{cross}}$ |
|---|---|---|---|---|
| 100 | 35.5 | 36.2 | 2.5 | 2.4 |
| 200 | 24.0 | 31.7 | 4.0 | 5.2 |
| 300 | 24.4 | 27.4 | 6.4 | 9.3 |
| 400 | 24.8 | 32.7 | 8.2 | 12.7 |
| 500 | 26.3 | 29.4 | 8.6 | 14.6 |
| 600 | 26.2 | 29.9 | 9.5 | 18.6 |
| 700 | 33.5 | 33.7 | 11.5 | 21.0 |

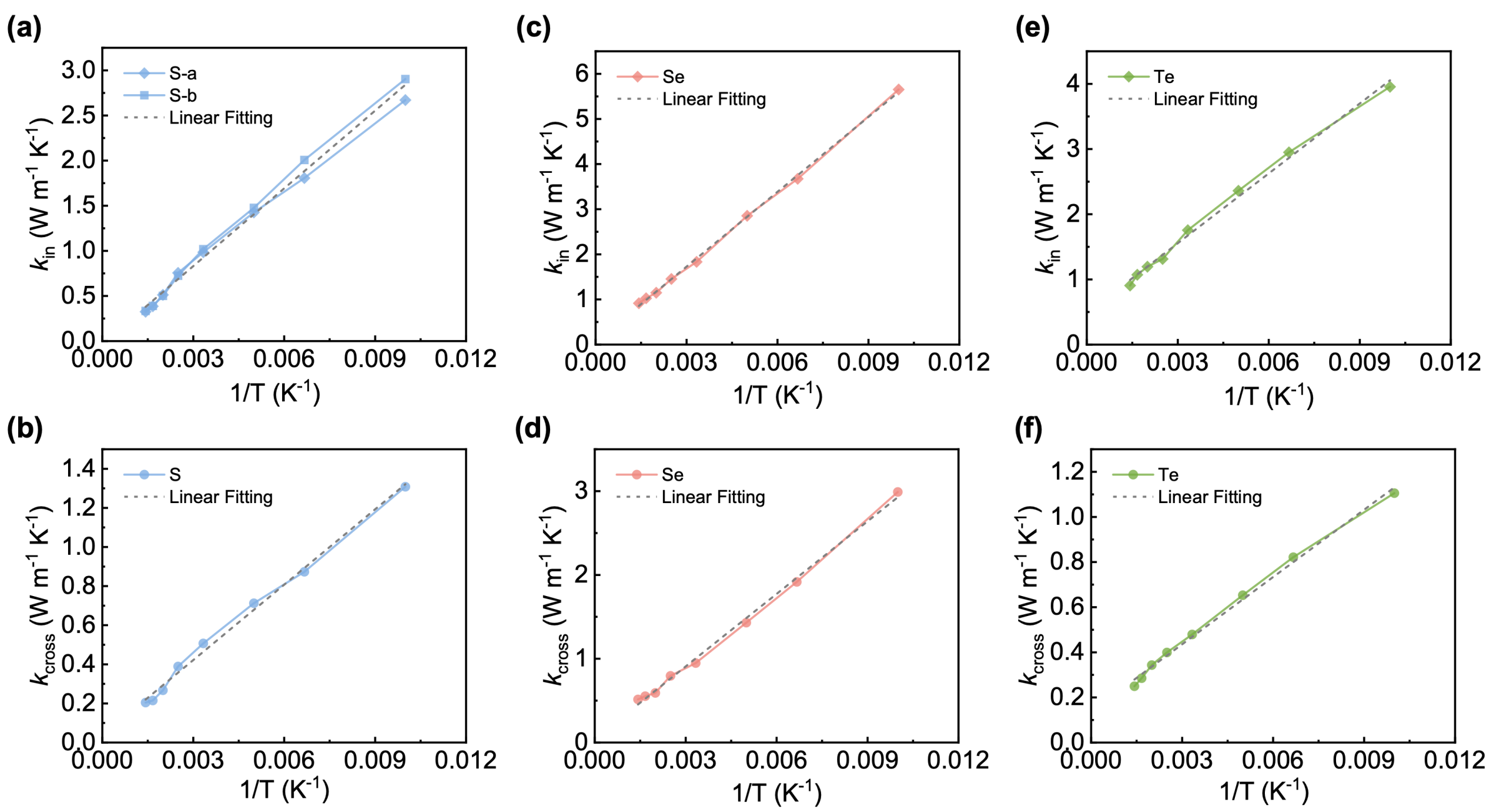

**Figure S6:** Lattice thermal conductivity of $Bi_2O_2S$, $Bi_2O_2Se$, and $Bi_2O_2Te$ replotted as a function of $1/T$. The top row shows the in-plane components and the bottom row shows the cross-plane components. Dashed lines are linear fits to the calculated data. The near-linear dependence indicates that $\kappa_L$ can be approximately described by an inverse relation with temperature over the investigated temperature range.

# VIII Device-scale simulations

The thermal meshes comprise the fin and substrate, whereas the thermoelastic meshes additionally include the gate dielectric. Their mesh-node counts are denoted by $N_{\mathrm{th}}$ and $N_{\mathrm{mech}}$, respectively. The thermal and thermoelastic meshes are linked by spatial mapping of the calculated temperature fields.

**Table S6:** Mesh-convergence results for the sequential GiftBTE–COMSOL simulations of $Bi_2O_2Se$ under the normal load ($Q_1 = 1.4 \times 10^{18}$ W m$^{-3}$) and elevated load ($Q_2 = 5.8 \times 10^{18}$ W m$^{-3}$). $T_{\mathrm{hot}}$ and $\varepsilon_{\mathrm{mis}}^{\mathrm{max}}$ denote the peak hotspot temperature and maximum thermal-mismatch strain, respectively.

| $N_{\mathrm{th}}$ | $N_{\mathrm{mech}}$ | Normal load, $Q_1$ | | Elevated load, $Q_2$ | |
|---|---|---|---|---|---|
| | | $T_{\mathrm{hot}}$ (K) | $\varepsilon_{\mathrm{mis}}^{\mathrm{max}}$ (%) | $T_{\mathrm{hot}}$ (K) | $\varepsilon_{\mathrm{mis}}^{\mathrm{max}}$ (%) |
| 16930 | 22370 | 346.2 | 0.66 | 514.0 | 1.82 |
| 32630 | 42466 | 346.3 | 0.65 | 515.1 | 1.83 |
| 65031 | 82951 | 346.7 | 0.68 | 514.7 | 1.85 |
| 125289 | 157457 | 346.8 | 0.68 | 516.1 | 1.85 |

**Table S7:** Peak hotspot temperatures $T_{\mathrm{hot}}$ (K) of $Bi_2O_2X$ fins under the elevated load ($Q_2 = 5.8 \times 10^{18}$ W m$^{-3}$), calculated using fixed phonon-property sets evaluated at $T_{\mathrm{prop}} = 300$, 500, and 700 K.

| Material | $T_{\mathrm{prop}} = 300$ K | $T_{\mathrm{prop}} = 500$ K | $T_{\mathrm{prop}} = 700$ K |
|---|---|---|---|
| $Bi_2O_2S$ | 659.43 | 755.65 | 826.91 |
| $Bi_2O_2Se$ | 516.14 | 616.73 | 700.64 |
| $Bi_2O_2Te$ | 567.09 | 631.35 | 751.23 |

**Table S8:** Material parameters adopted for the $HfO_2$ gate dielectric and $LaAlO_3$ substrate in the isotropic thermoelastic model. The listed values were treated as temperature-independent model parameters.

| Parameter | $HfO_2$ | $LaAlO_3$ | Sources |
|---|---|---|---|
| Density, $\rho$ (kg m$^{-3}$) | 9750 | 6524 | [12,13] |
| Young's modulus, $E$ (GPa) | 177 | 267.8 | [12,13] |
| Poisson's ratio, $\nu$ | 0.25 | 0.272 | [12,13] |
| Linear CTE, $\alpha$ ($10^{-6}$ K$^{-1}$) | 4.40 | 6.99 | [12,14] |

The $HfO_2$ Young's modulus and Poisson's ratio follow the ALD-film analysis of Gaskins et al. [12]; the reported in-plane CTE was adopted as an effective isotropic value. Its nominal density was selected within the reported 9.7–9.8 g cm$^{-3}$ range. For $LaAlO_3$, $E$ and $\nu$ were calculated from the Voigt–Reuss–Hill bulk and shear moduli, $K = 195.5$ GPa and $G = 105.3$ GPa [13], using the standard isotropic relations. The nominal density is consistent with the reported room-temperature value of 6.52 g cm$^{-3}$. The $LaAlO_3$ CTE was evaluated at 300 K as $\alpha = (3V)^{-1}(dV/dT)$ using the experimental volume fit of Chakoumakos et al. [14].

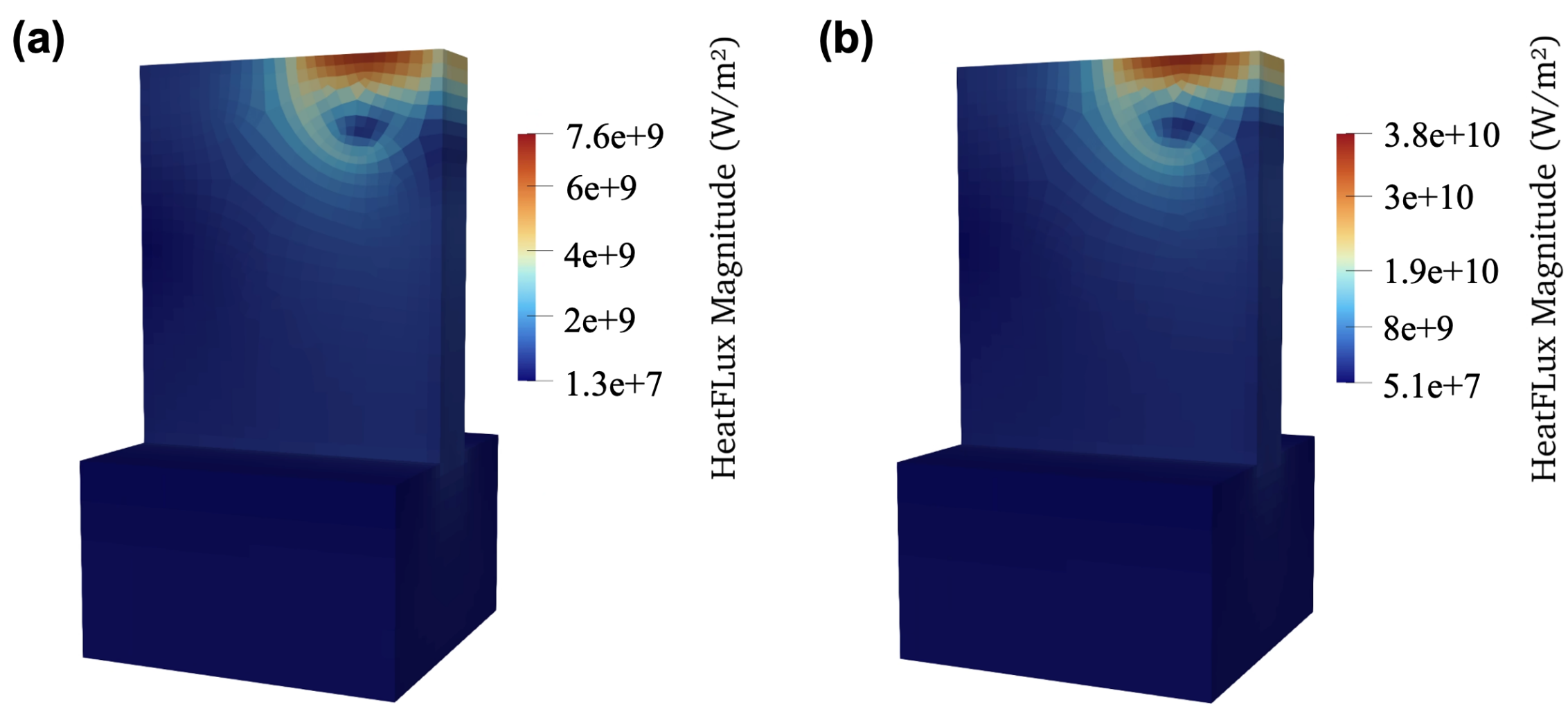

**Figure S7:** Steady-state heat-flux-magnitude distributions in the $Bi_2O_2Se$ fin-on-substrate structure under (a) normal-load and (b) elevated-load volumetric heat-generation rates of $1.4 \times 10^{18}$ W m$^{-3}$ and $5.8 \times 10^{18}$ W m$^{-3}$, respectively, obtained from the GiftBTE simulations.